\pdfoutput=1

\documentclass[11pt,twoside,a4paper,cmspaper,final,collab]{cms-tdr}
\def\svnVersion{270944}\def\cmsCernNo{2013-037}\def\cmsCernDate{\today}\def\cmsMessage{Published in the Journal of High Energy Physics as \href{http://dx.doi.org/10.1007/JHEP12(2014)017}{\doi{10.1007/JHEP12(2014)017.}}}

\begin{document}\cmsNoteHeader{JME-13-006}

\hyphenation{had-ron-i-za-tion}
\hyphenation{cal-or-i-me-ter}
\hyphenation{de-vices}

\RCS$Revision: 264096 $
\RCS$HeadURL: svn+ssh://alverson@svn.cern.ch/reps/tdr2/papers/JME-13-006/trunk/JME-13-006.tex $
\RCS$Id: JME-13-006.tex 264096 2014-10-15 20:52:33Z alverson $
\cmsNoteHeader{JME-13-006}
\title{Identification techniques for highly boosted W bosons that decay into hadrons}

\date{\today}

\abstract{
  In searches for new physics in the energy regime of the LHC, it is
  becoming increasingly important to distinguish single-jet objects that
  originate from the merging of the decay products of W bosons produced
  with high transverse momenta from jets initiated by single partons.
  Algorithms are defined to identify
  such W jets for different signals of interest, using techniques that
  are also applicable to other decays of bosons to hadrons that result in a single jet,
  such as those from highly boosted Z and Higgs bosons.
  The efficiency for
  tagging W jets is measured in data collected with the CMS detector at
  a center-of-mass energy of 8\TeV, corresponding to an integrated
  luminosity of 19.7\fbinv.  The performance of W tagging in data is
  compared with predictions from several Monte Carlo simulators.}

\hypersetup{%
pdfauthor={CMS Collaboration},%
pdftitle={Identification techniques for highly boosted W bosons that decay into hadrons},%
pdfsubject={CMS},%
pdfkeywords={CMS, physics, software, computing}}

\newcommand{\ifnpas}{\iffalse}
\newcommand{\ifpas}{\iftrue}

\maketitle 

\section{Introduction}
\label{sec:intro}

The LHC at CERN probes a new energy regime in particle physics, where searches for physics
beyond the standard model (SM) at high mass scale often involve objects with large transverse momenta (\pt).
In final states that contain the W$^{\pm}$ and Z gauge bosons or Higgs bosons (H),
it is possible to achieve a high selection efficiency through the use of hadronic decay channels.
At sufficiently large boost above order of $\pt>200$\GeV,
the final state hadrons from the W$\to \cPaq\cPq'$ decay merge into
a single jet, and the traditional analysis techniques relying on resolved jets
are no longer applicable.  However, in such cases the analysis of jet substructure
can be used to identify those jets arising from decays of W, Z or H bosons.
Because the values of the mass of the W and Z bosons are rather close to each other,
we do not distinguish the two, and refer to such jets
collectively as V jets, while the Higgs boson mass is significantly higher and can be distinguished.
The focus of this paper is solely on the identification of W jets, however, we note that many of the
procedures described are equally applicable for handling highly boosted Z and H
bosons.

Measurements of jet substructure observables related to identification of W bosons
have been previously reported by CMS~\cite{jetmass,topwtag_pas} and
ATLAS~\cite{atlasjetmass2,atlasjetmass}.
Several searches at CMS have employed jet substructure techniques for
identifying (``tagging") W jets and Z jets.  These include searches in
all-jet $\ttbar$ final states~\cite{hadronictop,hadronictop7TeV},
single and pair produced V bosons in inclusive dijet final
states~\cite{hadronicvv,hadronicvv7TeV}, and searches in the VV final states, where one
of the vector bosons decays
leptonically~\cite{semileptonicZZ,semileptonicZZ7TeV}.  In these
searches, a variety of different observables have been used to
identify the V jets.
This paper aims to compare and measure the performance in 8\TeV pp collisions of various jet substructure techniques that
can be used to distinguish V jets from more ordinary quark- and gluon-initiated jets,
which we refer to as QCD jets.

This paper is organized as follows.
The CMS detector is described in Section~\ref{sec:detector}.
The procedures chosen for the reconstruction of events are described in Section~\ref{sec:reco}.
The data and simulated events used in our studies as well as the event selection criteria are presented in Section~\ref{sec:samples}.
In Section~\ref{sec:algo}, through Monte Carlo (MC) simulation, we investigate the performance of jet substructure observables used to identify W jets, in order to find the best discriminants for such events.
We compare these observables in different kinematic regimes, and examine factors that contribute to their performance.
Their distributions in data are compared to those in MC simulations in Section~\ref{sec:efficiency}, to learn how well current MC simulations can model the physical processes responsible for jet substructure.
The methods used to extract data-to-simulation scale factors needed to correct W boson tagging efficiencies obtained from MC simulation are discussed in Section~\ref{sec:efficiency}, and the mistagging rate of QCD jets in data is extracted.
The goal being to provide these as reference tools for analyzing events with jets from V bosons in the final state.
Finally, we give a summary of our studies in Section~\ref{sec:summary}.

\section{CMS detector}
\label{sec:detector}

The central feature of the CMS detector is a 3.8\unit{T} superconducting
solenoid of 6\unit{m} internal diameter.
A complex silicon tracker, a crystal electromagnetic calorimeter (ECAL), and
a hadron calorimeter (HCAL) are located within the
magnetic field volume.  A muon system is
installed outside the solenoid, and embedded in the steel return yoke.
The CMS tracker consists of 1440 silicon pixel and 15\,148 silicon
strip detector modules.  The ECAL consists of 75\,848 lead
tungstate crystals, which provide coverage in pseudorapidity of
$\abs{\eta}< 1.48$ in the central barrel region and $1.48 <\abs{\eta}< 3.00$ in the two forward endcap regions.
The muon system
includes barrel drift tubes covering the pseudorapidity range $\abs{\eta}<
1.2$, endcap cathode strip chambers ($0.9< \abs{\eta}< 2.5$), and
resistive plate chambers ($\abs{\eta}< 1.6$).
A more detailed description of the CMS detector, together with a definition of the coordinate system used and the relevant kinematic variables, can be found in Ref.~\cite{Chatrchyan:2008aa}.

\section{Event reconstruction}
\label{sec:reco}

Jets are reconstructed by clustering particles obtained using the
particle flow (PF)
algorithm~\cite{CMS-PAS-JME-10-003,particleflow,particleflow2}.  The
PF procedure identifies each individual particle (a PF candidate)
through an optimized combination of all subdetector information.  The
energy of photons is obtained directly from the ECAL measurement,
corrected for suppression effects of energies from
calorimetric channels with small signals
(referred to as zero-suppression)~\cite{Chatrchyan:2013dga}.  The energy
of an electron is determined from a combination of the track momentum
at the main interaction vertex, the corresponding ECAL cluster energy,
and the energy sum of all bremsstrahlung photons associated with the
track. The energy of a muon is obtained from the corresponding track
momentum. The energy of a charged hadron is determined from a
combination of the track momentum and the corresponding ECAL and HCAL
energies, corrected for zero-suppression effects, and calibrated for
the nonlinear response of the calorimeters. Finally, the energy of a
neutral hadron is obtained from the calibrated energies in ECAL and
HCAL.

The PF candidates are clustered into jets using two algorithms: the
anti-\kt algorithm~\cite{antiktalg} with the distance parameter $R = 0.5$
(AK5), and the Cambridge-Aachen algorithm~\cite{CAaachen,CAcambridge}
with the distance parameter $R = 0.8$ (CA8), as implemented in \textsc{FastJet}
version 3.0.1 \cite{Cacciari:2011ma}.
While the CA8 algorithm with a larger distance parameter is used throughout
this paper to select and identify W jets, the AK5 algorithm is used to put
requirements on additional QCD jets in the event selection.
The choice of these algorithms is further explained in section~\ref{sec:algo}.
To mitigate the effect of multiple interactions in the same
bunch crossing, the so-called pileup (PU), charged
hadrons that are not associated with the primary vertex are removed from
the list of PF candidates.
The procedure is referred to as charged-hadron subtraction~\cite{JME-14-001} and strongly
reduces the dependence of the jet energy and substructure reconstruction on pileup.
An event-by-event jet-area-based
correction~\cite{jetarea_fastjet,jetarea_fastjet_pu,JME-JINST} is
applied to remove the remaining energy due to neutral particles
originating from the other pp collision vertices.
All jet substructure observables are computed using PF candidates
calibrated prior to jet clustering.  However, the resulting jets require
another small correction to the jet momentum and energy that
accounts for tracking inefficiencies and threshold effects.
The typical jet energy resolution is 5--10\% for jets with $\pt>200\GeV$.

Two algorithms are used to reconstruct
muons~\cite{CMS-PAPER-MUO-10-004}: one proceeds from the inner tracker
outwards, while the other starts from tracks measured in the muon
chambers and matches them to those reconstructed in the silicon
tracker. Muons are identified using selection criteria optimized
for high-\pt muons~\cite{CMS-PAPER-MUO-10-004}.  The selected muon
candidates must be isolated from charged hadron activity in the
detector by requiring the scaler sum of transverse momenta ($I_\mathrm{tk}$) of
tracks within a cone of
$\Delta R = \sqrt{(\Delta \phi)^2 + (\Delta \eta)^2} < 0.3$
around the muon track, divided by the
muon \pt, to be $I_\mathrm{tk}/\pt < 0.1$.  Electrons are reconstructed
using a Gaussian-sum filter algorithm~\cite{Chatrchyan:2013dga,CMS-PAS-EGM-10-004}, and
each electron candidate must furthermore pass the identification and
isolation criteria optimized for high \pt
electrons~\cite{CMS-PAS-EGM-10-004}.

\section{Data and simulated event samples}
\label{sec:samples}

\subsection{Event topologies}

This study aims to distinguish W jets from QCD jets.
We use three different final state topologies to establish
W jet identification in a broad region
of phase space, thereby enabling a number of physics data analyses.
In the $\ttbar$-enriched lepton+jets event topology,
the decay of two top quarks results in a final state with
two b quarks and two W bosons of which one decays leptonically
and the other decays to hadrons.
This topology provides a relatively pure source of W jets in data,
and is used to compare the efficiencies of W-tagging
in data and in simulation.
In contrast, the W+jet event topology, where the W boson decays leptonically, and
the inclusive dijet event topology are used as a source of QCD jets
to study their W-jet tagging properties in data and in simulation.
These are the benchmark scenarios for
searches, where the leading backgrounds are SM W+jets
and dijet production.
The W+jet sample accesses the low \pt regime, while the dijet sample
reaches higher \pt, and therefore both samples are explored.
To study the discrimination of W jets and QCD jets in the W+jet and dijet topologies,
we use simulated samples of beyond-SM resonances decaying to the
WW final state as source of W jets.

\subsection{Data and simulated event samples}

The data were collected with the CMS detector at a proton-proton (pp)
center-of-mass energy of 8\TeV and correspond to an integrated
luminosity of $19.7\pm0.5$\fbinv~\cite{LUM-13-001}.

As the default simulated signal sample, we consider a resonance X that decays to a pair of
longitudinally polarized W bosons.  Such samples are
produced by considering either a warped extra-dimensional model, where
the SM fields propagate in the
bulk~\cite{GravitonWWZZ1,GravitonWWZZ2,GravitonWWZZ3}, or models with
SM-like high mass H bosons. Graviton resonance samples in the
extra-dimensional model are produced
with the \textsc{JHUgen} 3.1.8~\cite{Gao:2010qx,Bolognesi:2012mm},
interfaced with \PYTHIA~6~\cite{Sjostrand:2006za} for parton
showering including the effect of hard gluon radiation.
\PYTHIA~6.426 is used with Tune Z2*~\cite{FWD-11-003} in this paper.
SM-like H boson samples are produced with {\POWHEG
1.0}~\cite{Nason:2004,Frixione:2007,Alioli:2010xd} interfaced with
\PYTHIA~6.
To study the effect of W boson polarization on the distributions of
substructure variables, the model with the SM Higgs-like couplings is
compared to a model with a purely pseudoscalar H boson which yields only
transversely polarized W bosons.  These samples are produced with the
\textsc{JHUgen} and \PYTHIA~6, with a resonance width of $\approx$1\% chosen to
be narrower than the experimental resolution of 5--10\%.

The background is modeled using QCD multijet, W+jets, WW/WZ/ZZ, Drell--Yan
($\cPq\cPaq\to\cPZ/{\gamma^{*} \rightarrow\ell\ell}$),
$\ttbar$, and single top quark MC simulation samples.
Three QCD multijet samples are compared.
A first sample is generated with \MADGRAPH~v5.1.3.30~\cite{madgraph}, with showering and hadronization
performed with \PYTHIA~6.
The second sample is generated as well as evolved with \HERWIG{++}~2.5.0~\cite{herwig}
with tune version 23~\cite{herwig}.
The third sample is generated with \PYTHIA~8.153~\cite{Sjostrand:2007gs} with Tune 4C.
\MADGRAPH, \PYTHIA~6 and \PYTHIA~8 are used with the CTEQ61L~\cite{cteq} parton distribution functions (PDF),
while \HERWIG{++} is used with the MRST2001~\cite{mrst} PDF.
Two W+jets samples with different parton shower models are compared:
one sample generated with \MADGRAPH interfaced with \PYTHIA~6 and
a second sample generated with \HERWIG{++}.
The single top quark and $\ttbar$ samples
are simulated with \POWHEG interfaced with \PYTHIA~6 using the CT10~\cite{ct10} PDF.
An alternative $\ttbar$ sample, generated with
\MCATNLO~\cite{Frixione:2002ik} and evolved with \HERWIG{++} using the CTEQ6M~\cite{cteq} PDF,
is also used for studies of systematic effects.
The Z+jets process is simulated with \MADGRAPH interfaced with \PYTHIA~6.
The VV production processes are simulated with \PYTHIA~6.

All generated
samples are processed through a
\GEANTfour-based~\cite{Agostinelli:2002hh} simulation of the CMS
detector.
An average of 22 supplementary interactions are added to the generated
events in order to match the additional particle production observed in data
from the large number of PU proton-proton interactions occurring per LHC bunch
crossing.

\subsection{Event selection}

The dijet and W+jet topologies are chosen to be in the kinematic
regime typically considered in searches for new
phenomena~\cite{hadronicvv,semileptonicZZ}.
In both topologies we
focus on the W-jet tagging properties of the highest \pt CA8 jet in
the event, requiring $\abs{\eta} < 2.4$, so that the core of the jet falls
within the tracker acceptance.  The ranges in jet \pt and the resonance
masses $m_\mathrm{X}$ are chosen to have the \pt distributions
similar for signal and for background.  For the W+jet topology, the
jet \pt is within 250--350\GeV and $m_\mathrm{X}$ = 600\GeV,
while for the dijet topology, the jet \pt is within 400--600\GeV and
$m_\mathrm{X}$ = 1\TeV.

Collision data events with a dijet final state are collected using the logical ``OR'' of a set of
triggers based on requirements on $H_{\mathrm{T}} = \sum_{\text{jets}}
\pt$ (scalar sum of $\pt$ of the AK5 jets), and on the
invariant mass of the two jets of highest $\pt$.  Subsequent event
selection follows closely the VV resonance search in
Ref.~\cite{hadronicvv}.  Events are initially selected by requiring at
least two jets with $\pt > 30\GeV$ and $\abs{\eta} < 2.4$.  The two jets
of highest $\pt$ are required to have a pseudorapidity separation
$\abs{\Delta\eta}<1.3$, which rejects a large fraction of QCD multijet
events.  Finally, the dijet invariant mass is required to be
larger than 890\GeV.  This threshold is chosen such that the trigger
selection for events with dijet masses above this threshold is 99\% efficient.
W-tagging is studied using the leading
jet in the selected dijet events, with additional requirements set on jet
$\pt$.

The main goal of the kinematic selection of the W+jet sample is to
isolate a sample of events with a highly boosted topology
consistent with a leptonically
decaying W boson recoiling against a high \pt jet.  The W+jet sample,
as well as the $\ttbar$ sample discussed below, are
collected using single-lepton triggers.  The lepton \pt thresholds of
these triggers are 40 and 80\GeV for the muon and electron channels,
respectively.  Offline, at least one muon or one electron, with respective $\pt >50$\GeV or
$\pt >90$\GeV, is required within respective $\abs{\eta}<2.1$ or $\abs{\eta}<2.5.$  Events
containing additional muons with $\pt > 20$\GeV and $\abs{\eta}<2.4$
or additional electrons with $\pt > 20$\GeV and $\abs{\eta}<2.5$ are rejected, in
order to improve the purity of W+jet events.  A requirement on the
imbalance in transverse momentum (\ETmiss) is used to reduce the QCD
multijet background.
The \ETmiss is computed from the negative
transverse component of the vector sum of all PF candidate momenta,
and is required to be above 50\GeV or 80\GeV for the
muon and electron channel.
The threshold is higher in the electron channel to further suppress
the larger background from multijet processes.
The \pt of the leptonically decaying W
boson and of the CA8 jet with highest \pt,
are required to be $>$200\GeV.  Additional criteria are applied to
ensure that the leptonic W boson and the CA8 jet are
mostly back-to-back in the transverse
plane: $\Delta R$ between the lepton and the jet must be greater than
$\pi/2$; the azimuthal distance $\Delta \phi$ between \ETmiss and the
jet must be greater than 2.0 radians; and the azimuthal distance
$\Delta \phi$ between the leptonically decaying W boson and the CA8 jet
must also be greater than 2.0 radians.
Finally, a cutoff on
additional jet activity in the event is applied to reduce the amount
of \ttbar\ background.
We identify additional b jet candidates in the event
by requiring that an AK5 jet,
with an angular distance of $\Delta R>0.8$ to the CA8 jet,
passes the CSV b-tagging discriminant~\cite{Chatrchyan:1494669}
using a medium working point.
To suppress \ttbar\ background in the W+jet selections described above,
we require that no such b jets be present in the event.

To select the $\ttbar$ sample, we use the kinematic
selection described above for the W+jet topology, but
instead require that there is at least one AK5 b jet,
with an angular distance of $\Delta R>0.8$ to
the CA8 jet considered as W jet candidate.
To increase the statistical precision of the sample, we
select the CA8 jet with the largest mass and with $\Delta\phi$ between
the lepton and the jet greater than $\pi/2$ as W jet candidate,
rather than the highest \pt CA8 jet.

\section{Algorithms for W jet identification}
\label{sec:algo}

A jet clustering algorithm with $R = 0.8$ is used to identify W jets.
A large value of $R$ increases the efficiency to reconstruct W bosons
with small boost as single jets, since the average angular distance
between the W decay products is inversely proportional to the
\pt of the W.
The chosen value of $R$ provides a high efficiency
for W bosons with small boost and ensures that no efficiency is lost
in the transition from classical W reconstruction from two small jets at low W \pt and
reconstruction from a single large jet at higher W \pt (see \eg Ref.~\cite{Gouzevitch:2013qca}).
Another point to consider when choosing the value of $R$, is the
$\ttbar$ data sample available for validating
highly boosted W jets.
If $R$ is chosen too large, the b quark from the $\cPqt\to\PW\cPqb$ decay tends to merge into the W jet.
The chosen value of $R$ is the result of a compromise between high efficiency
for W bosons with small boost and a sufficiently large sample of
W jets in $\ttbar$ data for validating the W jet identification algorithms.

\begin{figure}[th!b]
\begin{center}
\includegraphics[width=0.5\textwidth]{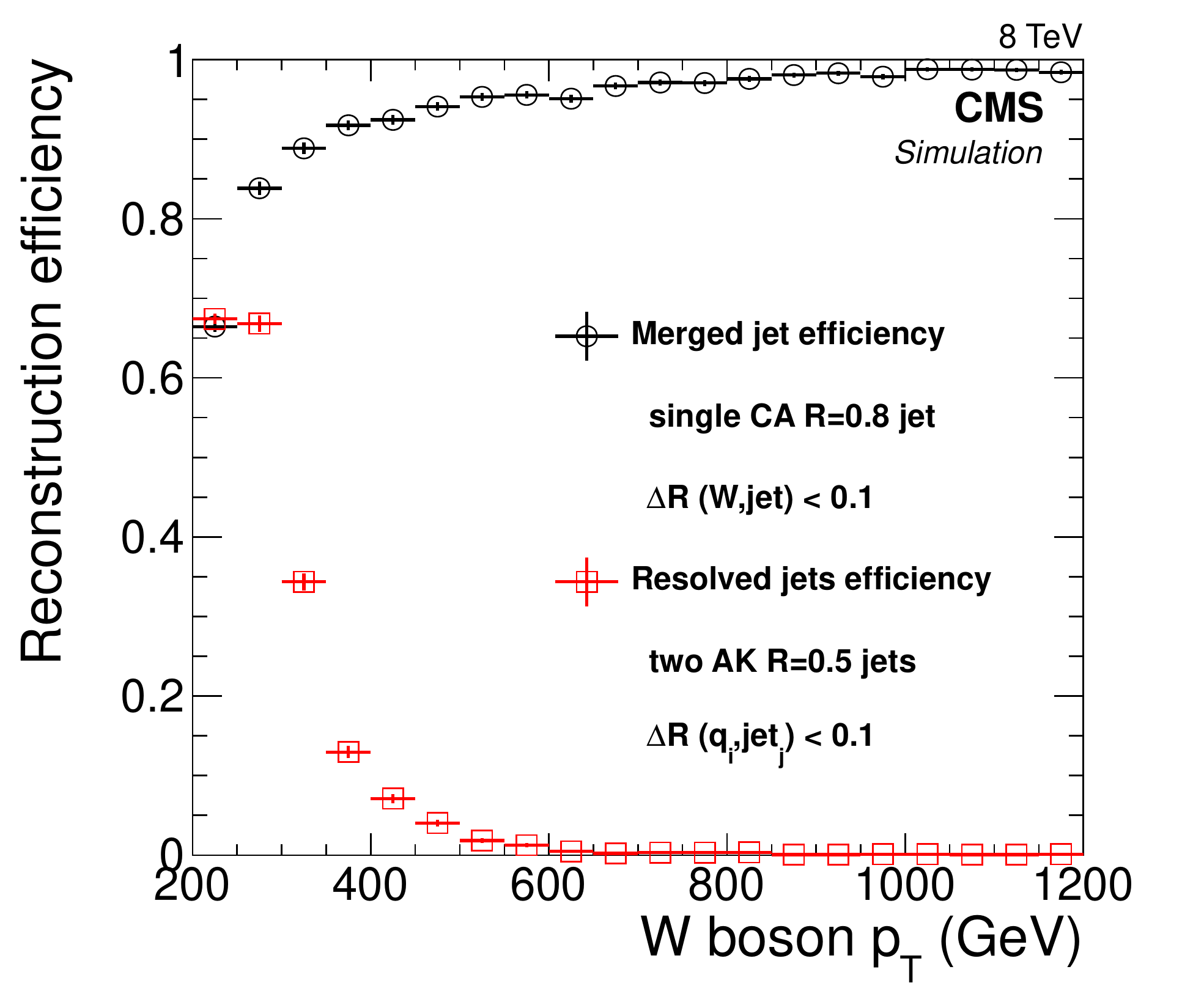}
\end{center}
\caption{Efficiency to reconstruct a CA8 jet within $\Delta R<0.1$ of a generated W boson, and the
efficiency to reconstruct two AK5 jets within $\Delta R<0.1$ of the generated quarks from
longitudinally polarized W bosons, as a function of the \pt of the W boson.}
\label{fig:ca8ak5}
\end{figure}
Figure~\ref{fig:ca8ak5} shows the \pt range of W bosons for which the
$R = 0.8$ algorithm is efficient and compares this to the efficiency for
reconstructing W bosons from two $R = 0.5$ jets.  Above a \pt of
200\GeV, the CA8 jet algorithm, used to identify W jets,
becomes more efficient than the reconstruction of a W boson from two AK5
jets.
In this paper we therefore study substructure observables to
identify W jets for an $R = 0.8$ algorithm.
Whether an AK or a CA algorithm is used in such comparison does
not affect the overall conclusion.  The choice of CA (with $R = 0.8$) and AK
($R = 0.5$) is simply due to their wide use in CMS publications,
where CA was introduced in the first top tagging algorithm paper of CMS~\cite{cmstop}.
Whenever we refer to efficiency ($\epsilon$) in this paper, we refer
to the full efficiency to identify a W boson relative to all generated
W bosons decaying to hadrons.

\subsection{Substructure observables}
\label{sec:algo_obs}

As the mass of the W boson is larger than the mass of a typical QCD jet,
the jet mass is the primary observable that distinguishes a W jet from
a QCD jet.  The bulk of the W jet mass arises from the kinematics of
the two jet cores that correspond to the two decay quarks.  In contrast,
the QCD jet mass arises mostly from soft gluon
radiation.  For this reason, the use of jet grooming methods such as
filtering~\cite{Butterworth:2008iy}, trimming~\cite{Krohn:2009th}, or
pruning~\cite{jetpruning1,jetpruning2}, improves discrimination by
removing the softer radiation, as this shifts the jet mass of QCD jets to
smaller values, while maintaining the jet mass for W jets close to the
W mass.  Studies of these grooming methods have been performed in
Ref.~\cite{jetmass}, with the conclusion that the pruned jet mass
provides the best separation between W signal and QCD background.  In
this paper, we use the grooming parameters proposed by the original authors.

\begin{figure*}[th!b]
\centering
\includegraphics[width=0.32\textwidth]{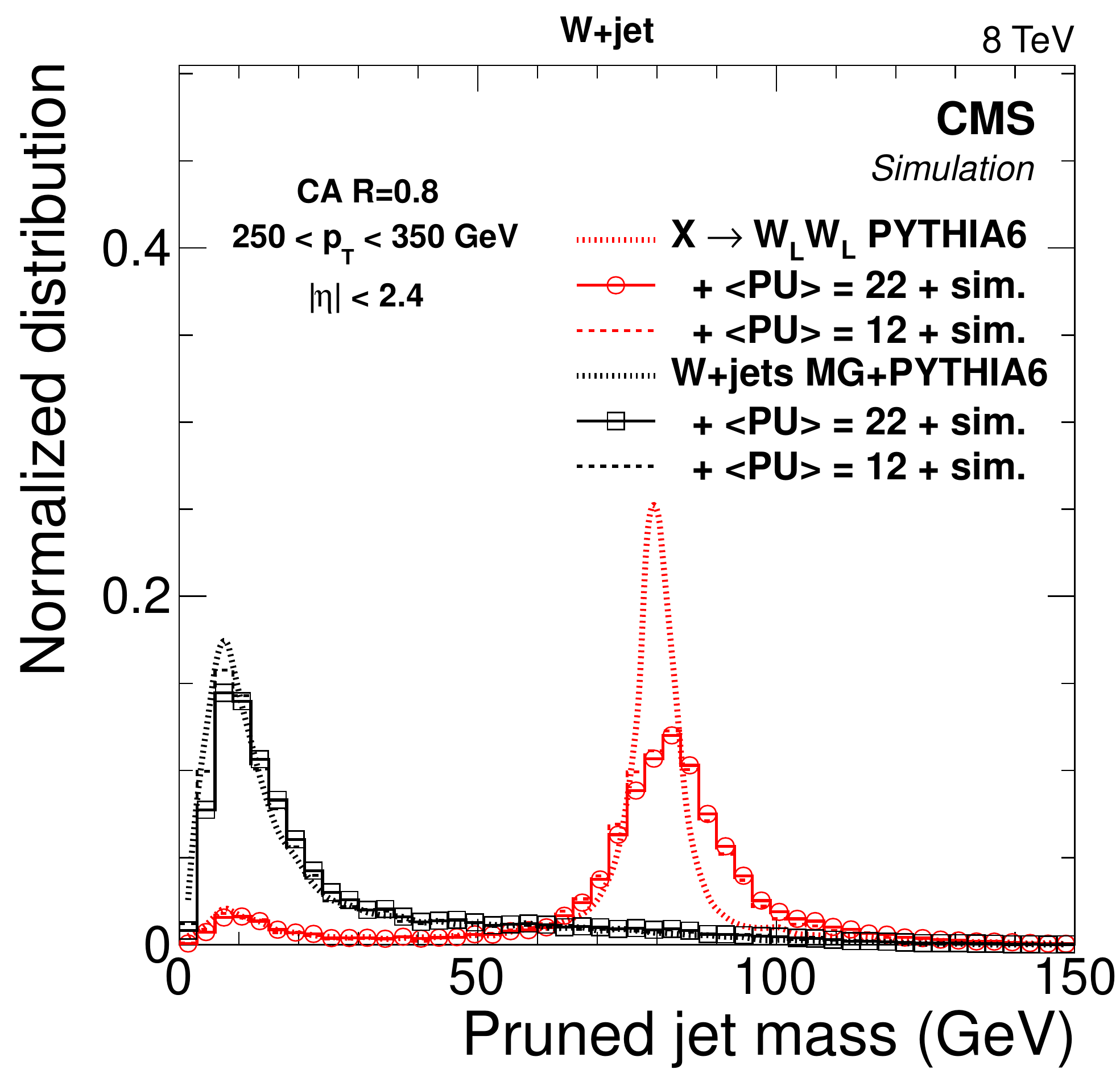} \includegraphics[width=0.32\textwidth]{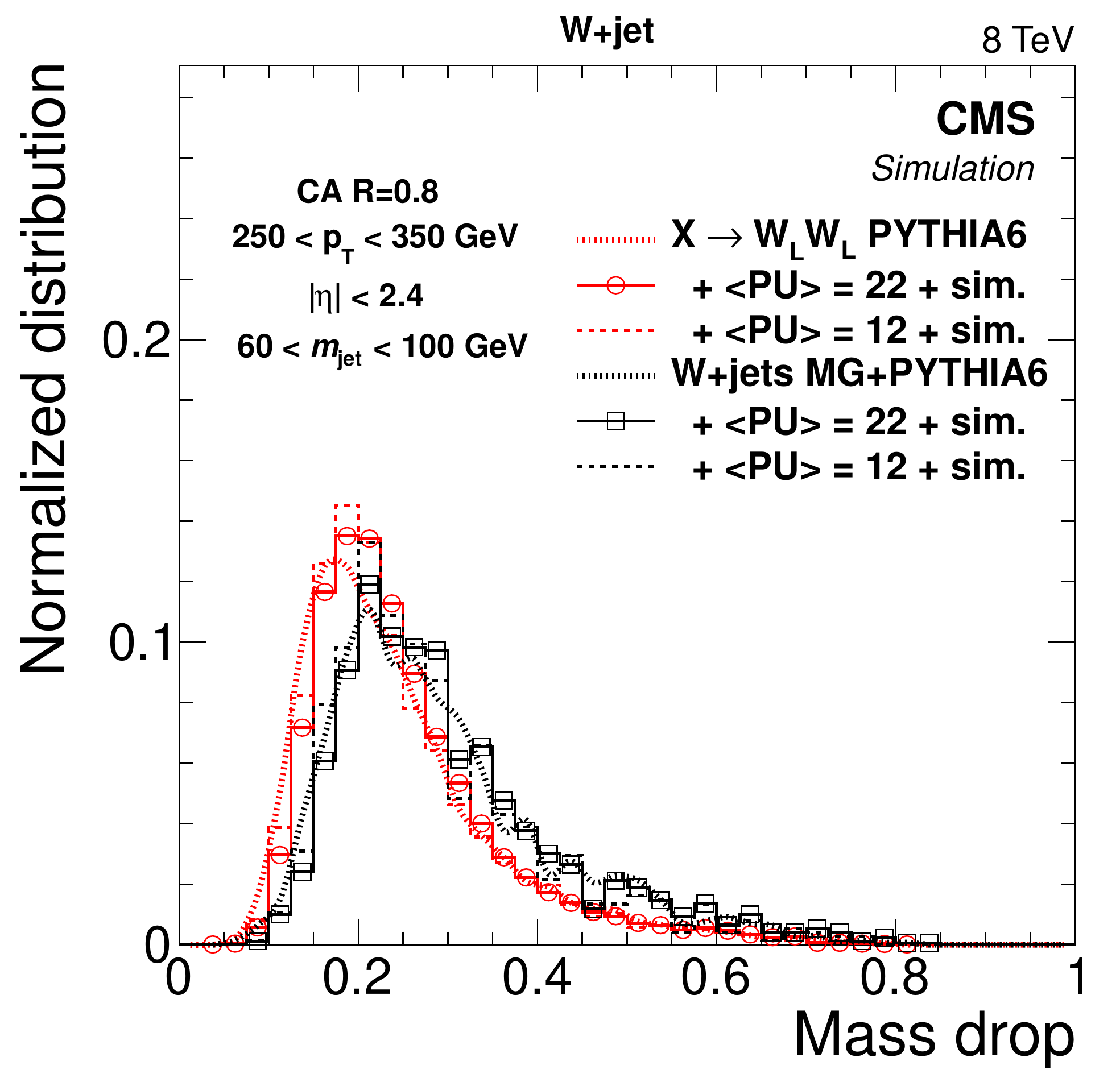} \includegraphics[width=0.32\textwidth]{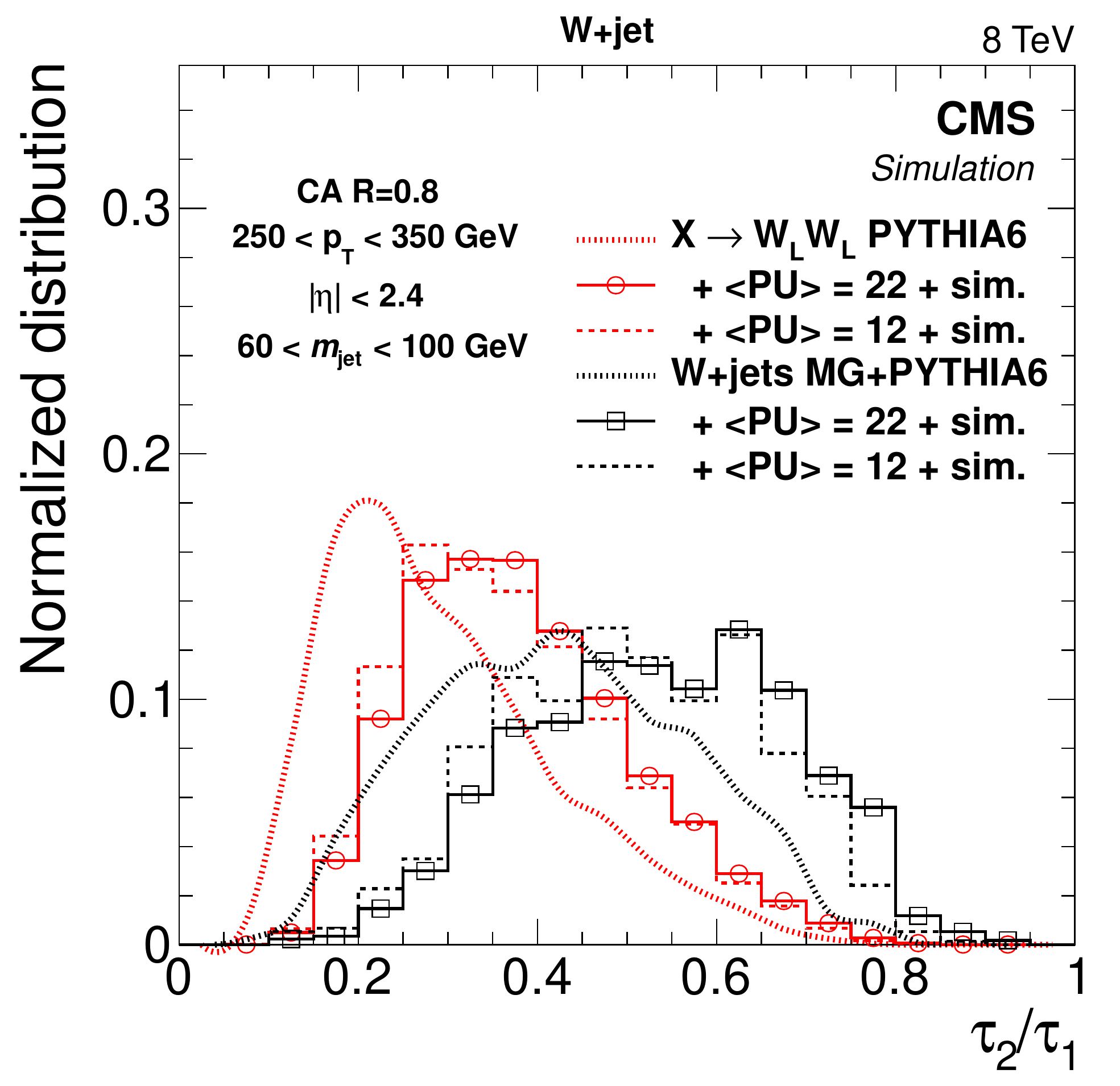} \includegraphics[width=0.32\textwidth]{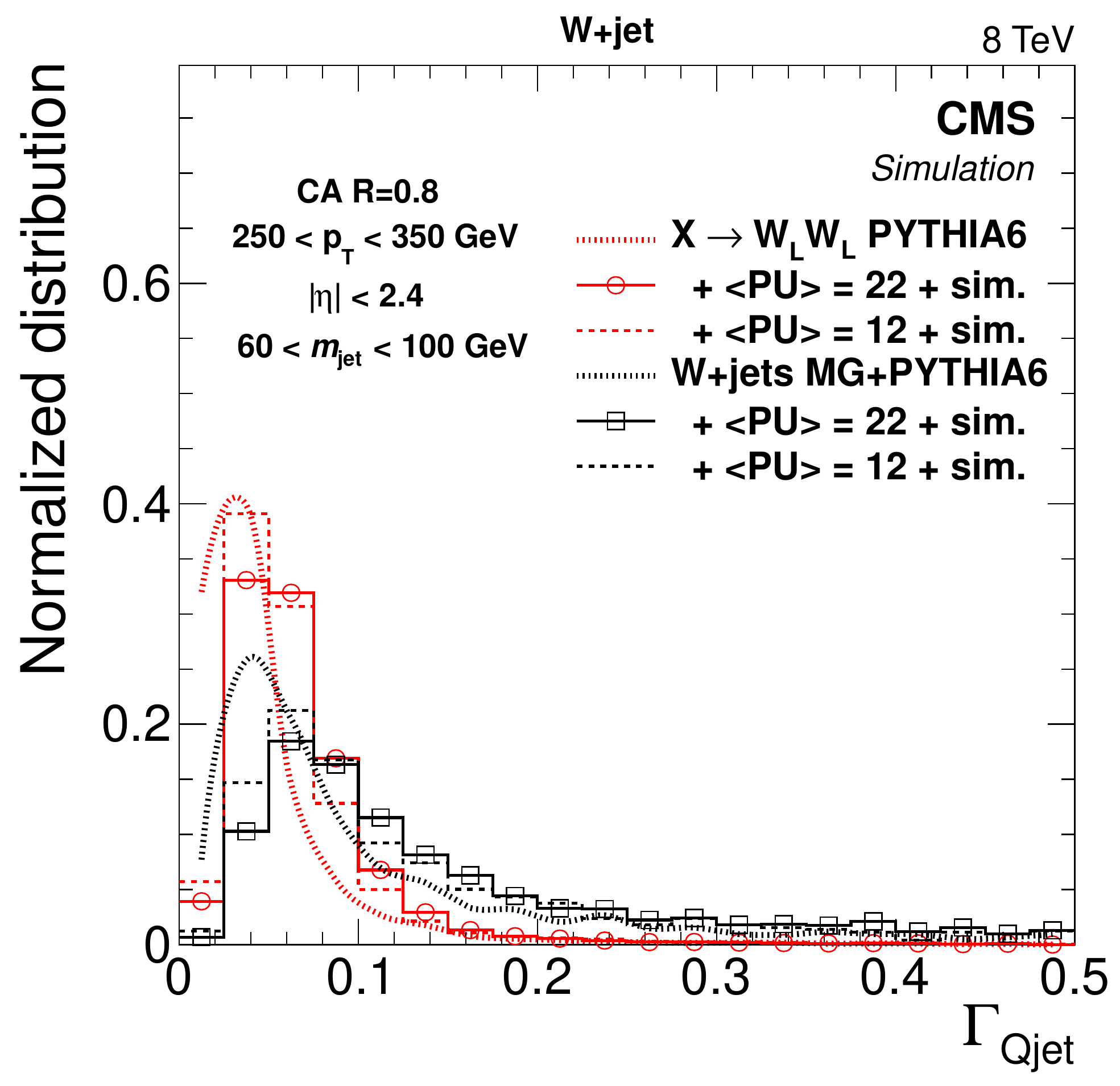} \includegraphics[width=0.32\textwidth]{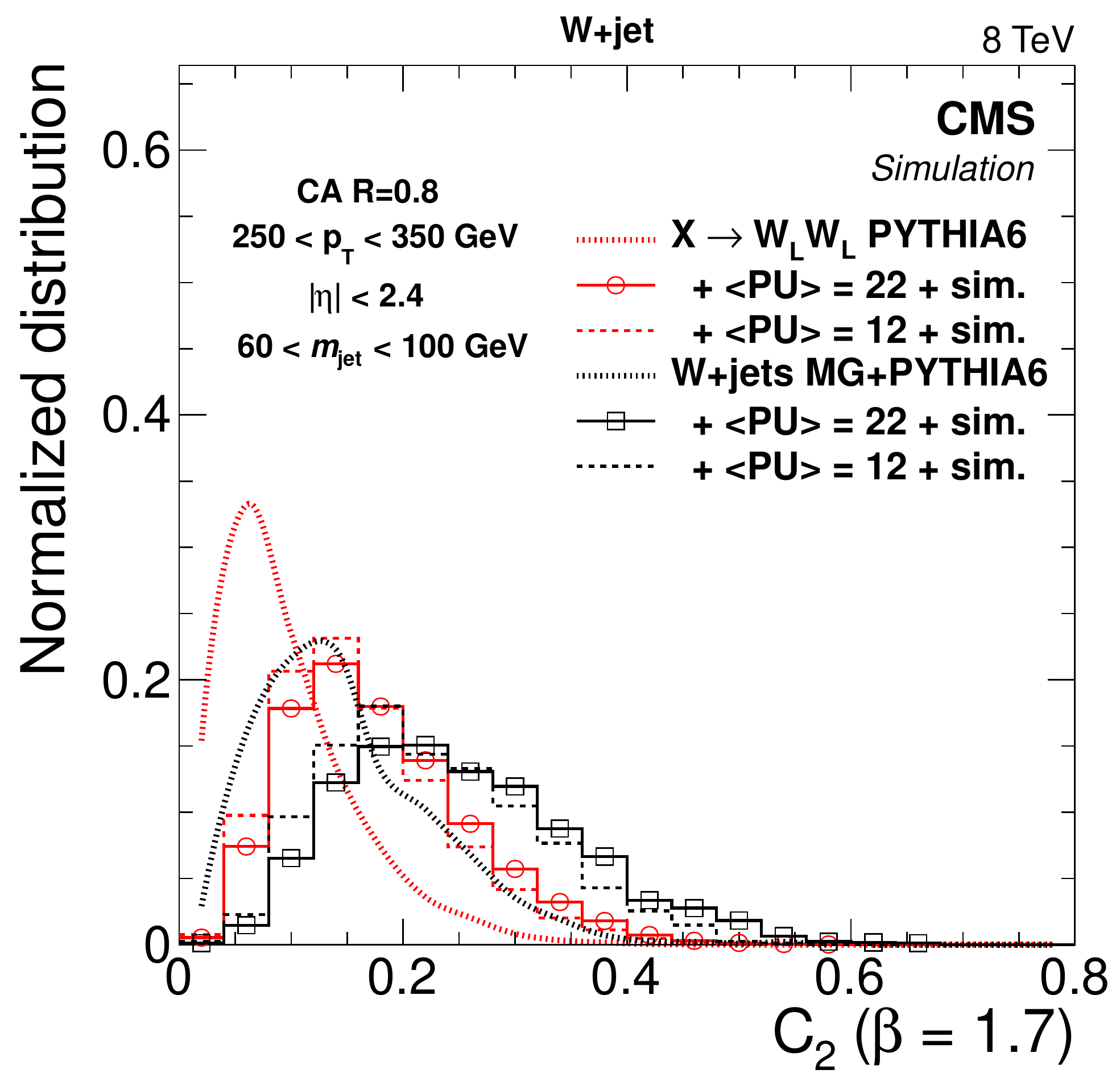} \includegraphics[width=0.32\textwidth]{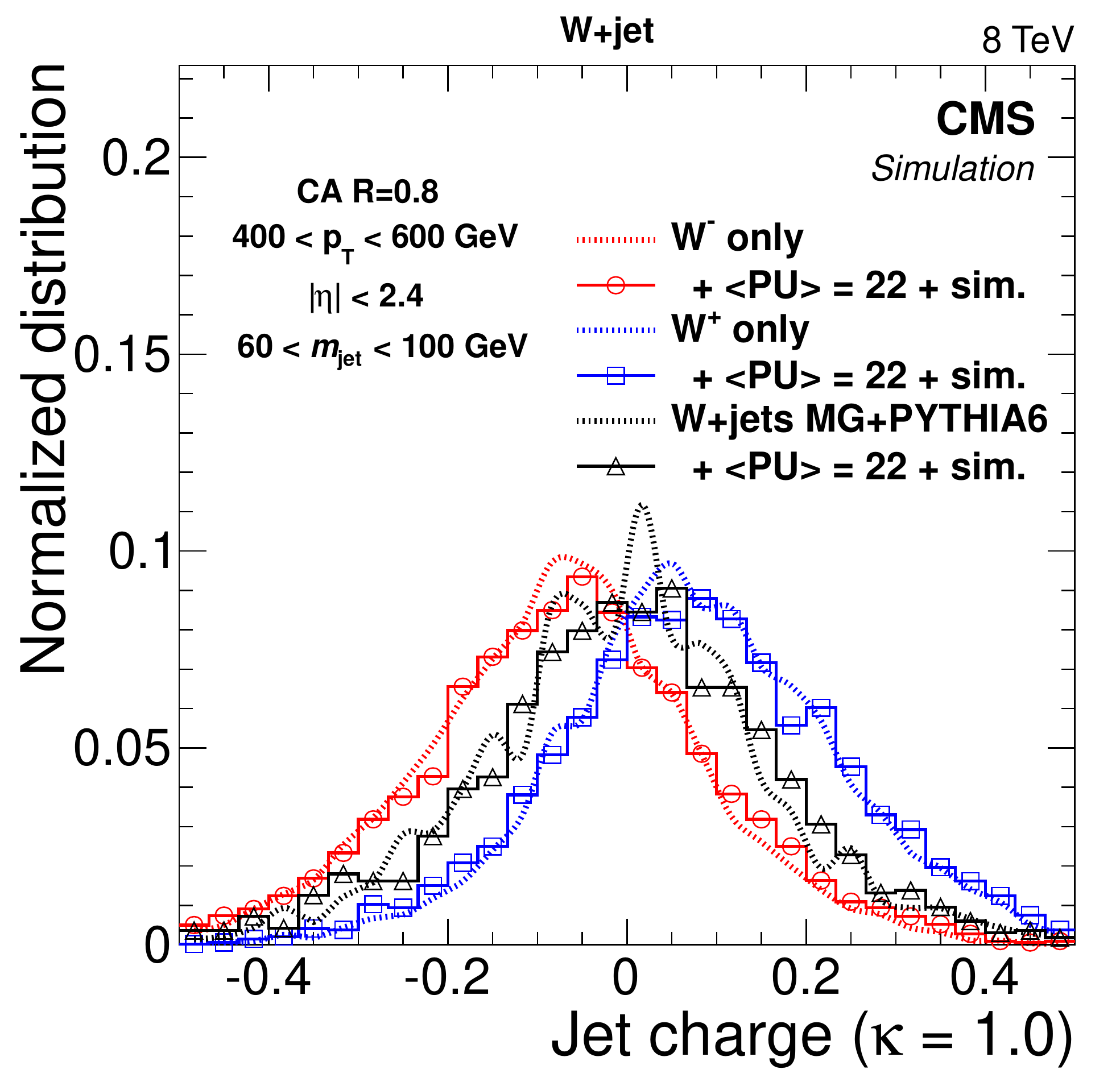}
 \caption{
Distributions of six variables characterising jet substructure in simulated samples
of highly boosted and longitudinally polarized W bosons and inclusive
QCD jets expected in the W+jet topology.  The discriminator
distributions (except for the pruned jet mass in the upper left panel) are shown after a selection on the
pruned jet mass of $60 < m_{\text{jet}} < 100\GeV$.
MG denotes the \MADGRAPH generator.  Thick
dashed lines represent the generator predictions without pileup
interactions and without CMS detector simulation.  The histograms are the
expected distributions after full CMS simulation with pileup
corresponding to an average number of 12 and 22 interactions.
(upper middle) gives the mass drop variable,
(upper right)  the N-subjettiness ratio $\tau_2/\tau_1$,
(lower left) the Qjet volatility,
(lower middle) the energy correlation function double ratio $C_{2}^\beta$,
and (lower left) the jet charge.}
\label{fig:algorithms_jetmass}
\end{figure*}

\textbf{Pruned jet mass} is obtained by removing the softest components
of a jet.  The CA8 jet is reclustered from its original jet
constituents, however the CA clustering sequence is modified to
remove soft and wide-angle protojets (single particles, or groups
of particles already combined in the previous steps).
In each recombination step, its hardness $z$ is defined as
$z=\min\{\pt^i,\pt^j\}/\pt^p$, where $\pt^i$ and $\pt^j$ are the $\pt$
of the two protojets to be combined and $\pt^p$ is the $\pt$ of
the combination of the two protojets.
The protojet with the lower $\pt^i$ is ignored if
$z < z_{\text{cut}}=0.1$, and if it forms an angle $\Delta R$ wider
than $D_{\text{cut}}=m^\text{orig}/\pt^\text{orig}$ relative to the axis of the
combination of the two protojets, where $m^\text{orig}$ and $\pt^\text{orig}$ are the
mass and \pt of the original CA8 jet.
The pruned jet mass distributions for W jets and QCD jets are shown
in Fig.~\ref{fig:algorithms_jetmass}\,(upper left) at generator level and
detector level with pileup.
Comparing the generator level predictions for the pruned jet mass of W jets
with those at detector level with pileup, the widening of the peak due to
detector resolution can be observed.

Further discrimination between W and QCD jets can be obtained from a more
extensive use of jet substructure.  Here we consider the following observables.

\textbf{Mass drop $\mu$}~\cite{Butterworth:2008iy} is calculated from the
two subjets that are obtained by undoing the last iteration of the CA jet
clustering via pruning.  The idea behind the mass drop is that the W
jet is formed by merging the showers of two decay quarks, and thus
the mass of each quark subjet is much smaller than the mass of the
W jet.  In contrast, a massive QCD jet is formed through continuous
soft radiation; the subjet with larger mass contains the bulk of
the jet and the ratio of the mass of the large subjet to the total
mass is therefore close to unity.
We define the mass
drop $\mu$ as the ratio of the masses of the higher mass subjet
($m_1$) and the total pruned jet ($m_{\text{jet}}$).
The two subjets can also be used to estimate their $\Delta R$,
which can provide additional discrimination.
The distribution of $\mu$ is shown in Fig.~\ref{fig:algorithms_jetmass}\,(upper middle).
The differences between the generator level predictions and those
at detector level with pileup are small for this observable, because
the detector can resolve the two relatively well separated subjets.

\textbf{N-subjettiness $\tau_N$}~\cite{Thaler:2010tr} is a generalized jet shape observable.
N-subjettiness is computed under the assumption that the jet has N subjets,
and it is the \pt-weighted $\Delta R$ distance between each jet constituent and
its nearest subjet axis:
\begin{equation}
\tau_N = \frac{1}{d_0} \sum_k p_{\mathrm{T},k} \min \{ \Delta R_{1,k},\Delta R_{2,k},\cdots, \Delta R_{N,k} \}
\end{equation}
where $k$ runs over all constituent particles.  The normalization
factor is $d_0 = \sum_k p_{\mathrm{T},k}R_0$ and $R_0$ is the original jet
distance parameter.  The $\tau_N$ observable has a small value if the
jet is consistent with having N or fewer subjets, as almost every
jet constituent will be close in $\Delta R$ to its own true subjet.
For discrimination between W jets with two subjets and QCD jets
consistent with corresponding to a single subjet,
the ratio $\tau_2/\tau_1$ is particularly
useful as it tends to smaller values for W jets.  The subjet
axes are obtained by running the exclusive \kt algorithm~\cite{ktalg},
and reversing the last N clustering steps.  The axes can be optimized
to minimize the N-subjettiness value.  As default, we use a ``one-pass"
optimization of the exclusive \kt axes, where one step of the
iterative optimization is performed.  By default $\tau_2/\tau_1$ is
calculated from the unpruned CA8 jets, but we also consider a pruned
$\tau_2/\tau_1$ calculated from pruned CA8 jets.
Fig.~\ref{fig:algorithms_jetmass}\,(upper right) shows the
$\tau_2/\tau_1$ distribution for W jets and QCD jets after requiring
$60 < m_{\text{jet}} < 100\GeV$, and demonstrates its
discrimination power after the pruned jet mass selection.
The distributions at detector level with pileup are shifted significantly
compared to the generator level predictions, though the discrimination
power is preserved. The shift was due equally to detector
effects and pileup.

\textbf{Qjet volatility $\Gamma_\text{Qjet}$}~\cite{Ellis:2012sn} is
a statistical measure of an ensemble of similar jet
clustering sequences.  A jet is defined by its cluster
sequence, which is topologically a tree and is here referred to as ``jet tree''.
By randomizing the recombination scheme and running the pruning algorithm
for each jet tree, we can define a family of trees for each jet from which
we can compute a distribution of jet masses.
The continuous soft radiation that forms massive QCD jets
results in clustering sequences susceptible to
fluctuations---a small deviation in soft radiation can result
in a very different order of putting the jet together.  In contrast,
W jets are characterized by two strong jet cores, and small
perturbations usually yield nearly identical clustering sequences.
Therefore a large volatility of the clustering sequence is
a characteristic of QCD jets, and can be used to distinguish
them from signal W jets.

The procedure for quantifying the volatility of the jet clustering
sequence is as follows.  At every step of clustering, a weight $w_{ij}$ is
assigned to each constituent pair, and then one of the
available pairs are randomly chosen and combined. The default weight
is defined as:
\begin{equation}
w_{ij} = {\rm exp}\{ -\alpha \frac{ d_{ij} - d^{\rm min} }{d^{\rm min}} \} ,
\end{equation}
where $d_{ij}= \Delta R^2_{ij}$ is the $(\eta,\phi)$ distance measure
of the CA algorithm within the $ij$ pair, $d^\text{min}$ is its minimum over
all pairs at this stage in the clustering, and $\alpha$ is the rigidity
controlling the level of randomness, where for $\alpha \to \infty$
represents the limit of a classical jet algorithm.
We choose to generate 50 random jet trees.  Qjet
volatility is defined as the root-mean-square (\textsc{rms}) of the jet mass distribution, divided
by the average jet mass, or $\Gamma_\text{Qjet} = \textsc{rms}/\langle
m \rangle$.  To improve the speed of the algorithm without greatly degrading
the performance, before Qjet clustering we pre-cluster the jet
constituents down to 35 protojets.
Fig.~\ref{fig:algorithms_jetmass}\,(lower left) shows the distributions in
$\Gamma_\text{Qjet}$.

\textbf{Energy correlation function double ratio \boldmath $C_{2}^\beta$}~\cite{ecf}
is defined as follows:
\begin{equation}
C^{\beta}_{2} = \frac{\sum_{i<j<k} p_{\mathrm{T}i} p_{\mathrm{T}j} p_{\mathrm{T}k} (R_{ij} R_{ik} R_{jk})^{\beta} \sum_{i} p_{\mathrm{T}i}}{(\sum_{i<j} p_{\mathrm{T}i} p_{\mathrm{T}j} (R_{ij})^{\beta})^{2}}
\end{equation}
where $i$, $j$ and $k$ runs over all constituent particles satisfying $i<j<k$.
Similarly to the ratio $\tau_2/\tau_1$, the numerator quantifies how likely a jet is composed of
two subjets, while the denominator gives a probability for being composed of one subjet.
We study $C^{\beta}_{2}$ with $\beta = 1.7$ as suggested in Ref.~\cite{ecf},
which is suited to discriminate two-prong
W jets from QCD jets consistent with having a single subjet.
The distribution of $C^{\beta}_{2}$ is given in Fig.~\ref{fig:algorithms_jetmass}\,(lower middle).

\textbf{Planar flow} with $R = 0.5$ and \textbf{trimmed grooming
sensitivity}~\cite{Cui:2010km} have also been considered in this
study.  Planar flow characterises the geometric distribution of energy
deposition from a jet, which discriminates W jets from QCD jets, as
the latter are more isotropic.  Trimmed grooming sensitivity is
defined as the decrease in jet mass, when the trimming
algorithm~\cite{Krohn:2009th} is applied to the jet.

\textbf{Jet charge, \boldmath $Q^{\kappa}$}~\cite{jetcharge} is
a measure of the electric charge of the parton that is the
origin of the jet.  This variable has a long history in flavor
tagging of neutral B mesons, and it is defined as
the \pt-weighted average charge of the jet:
\begin{equation}
\label{jetcharge_formula}
Q^{\kappa} = \frac{\sum_{i} q_i (\pt^{i})^{\kappa}}{(\pt^\text{jet})^{\kappa}}
\end{equation}
Here $i$ runs over all particles in a jet.  Our default choice for
$\kappa$ is 1.  It can be used to provide additional discrimination
among quark jets, gluon jets and W jets or also to distinguish
the charged W' signal from that of a neutral Z'. The differences between
the jet charge distribution of W$^\pm$ jets and of neutral jets can
be seen in Fig.~\ref{fig:algorithms_jetmass}\,(lower right).
Detector resolution and pileup have almost no effect on this variable
as it is built from charged hadrons identified using
the tracker where those from PU vertices are discarded.

\subsection{Comparison of algorithms}
\label{sec:opt}

We compare the performance of observables used to
identify W jets with the goal of establishing which
provides the best signal-to-background discrimination between W jets
and QCD jets.  Because the pruned jet mass is the best
discriminant, we examine the other variables only
for jets satisfying $60 < m_{\text{jet}} < 100\GeV$.
Observables highly correlated with the pruned
jet mass will therefore show weaker additional improvement in performance.

\begin{figure}[!htbp]
\begin{center}
\includegraphics[width=0.47\textwidth]{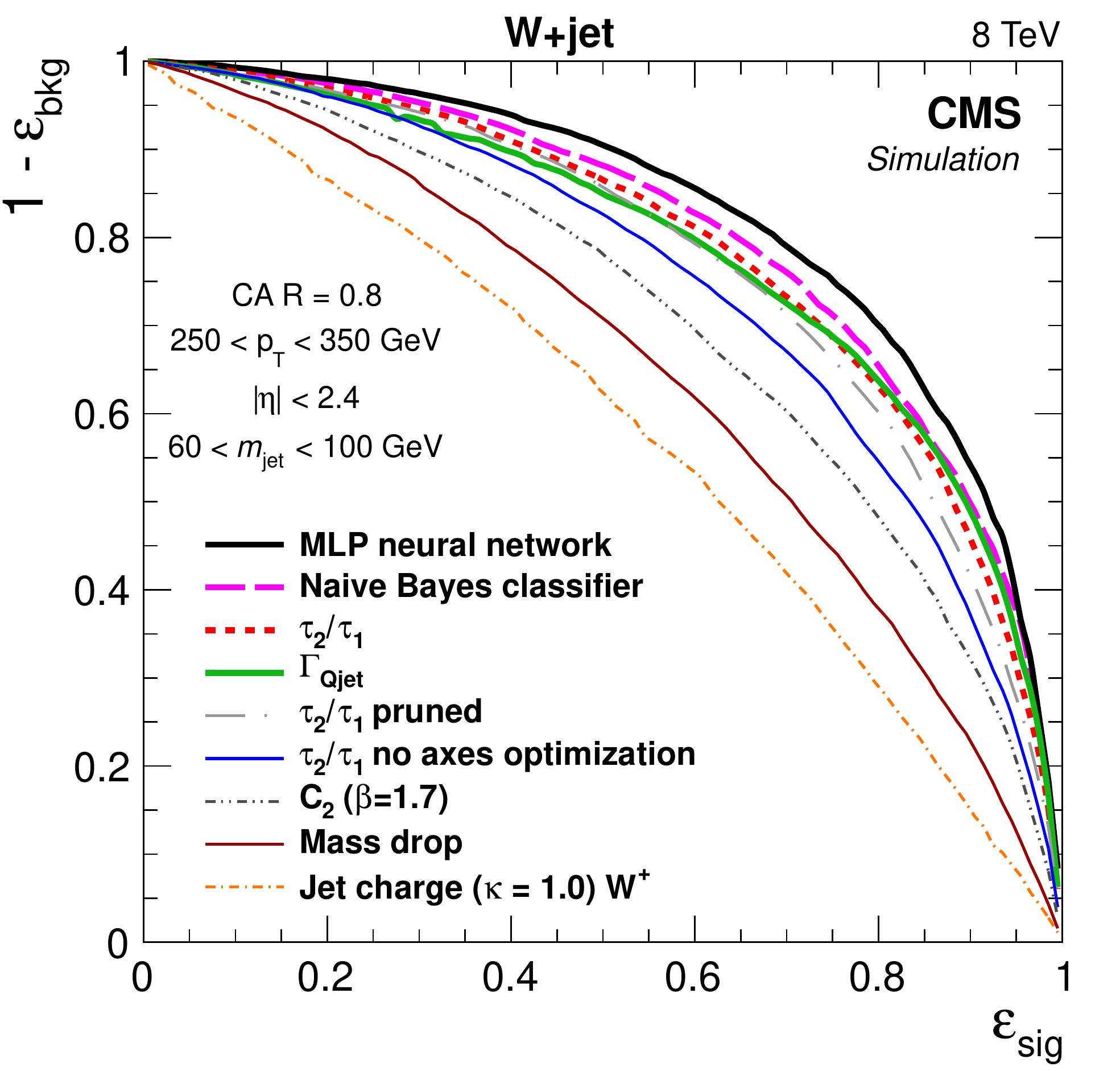}
\end{center}
\caption{Performance of several discriminants in the background-signal efficiency plane in the low jet \pt bin of 250--350\GeV in the W+jet topology.
The efficiencies and mistagging rates of the various discriminants are estimated on samples of W jets and QCD jets
that satisfy a pruned jet mass selection of $60 < m_{\text{jet}} < 100\GeV$.}
\label{fig:rocMVA}
\end{figure}

The figure of merit for comparing different substructure
observables is the background rejection efficiency as a function of
signal efficiency (``receiver operating characteristic'', or the ROC
curve).  Figure~\ref{fig:rocMVA} shows the performance of the observables
in the W+jet final state for jet \pt 250--350\GeV.  The
pruned jet mass selection is applied in both the numerator and the
denominator of the efficiency, and only the additional discrimination
power of the other observables is therefore shown in the figure.
The performance of the $\tau_2/\tau_1$, pruned $\tau_2/\tau_1$,
exclusive-\kt $\tau_2/\tau_1$, $\Gamma_\text{Qjet}$, $C_2^\beta$, mass
drop, and jet charge are compared.  For the jet charge ROC curve, a
positively charged lepton is required in the event selection, and
therefore the discrimination power of negatively charged W jets
against QCD jets is compared.  We find that the best performant
variable is $\tau_2 / \tau_1$ up to an efficiency of 75\%.  Above
an efficiency of 75\%, $\Gamma_\text{Qjet}$ is the best
variable.  The pruned $\tau_2 / \tau_1$ is slightly worse
than the default $\tau_2 / \tau_1$.  The performance of the $\tau_2
/ \tau_1$ without optimization of the axes is worse than the $\tau_2
/ \tau_1$ variants with a "one-pass" optimization.
The worst performing variables are the mass drop,
$C_2 (\beta = 1.7)$, and the
jet charge.  We also find that the discrimination power between $W^+$
jets and $W^-$ jets varies by less than 10\% for values of the
$\kappa$ parameter in Eq.~\ref{jetcharge_formula} between 0.3 and 1.0.

In addition to the performance of individual variables, we study how
their combination can improve the separation between W and QCD jets.
A multivariate optimization is performed using the {TMVA}
package~\cite{Hocker:2007ht}.
A combination of observables is
considered in a naive Bayes classifier and in a Multilayer Perceptron
(MLP) neural network discriminant.
Additional observables with respect to those shown in
Fig.~\ref{fig:rocMVA} are used in an attempt to
increase the discrimination power.
The variables used in both
discriminants are the mass drop, $\Gamma_\text{Qjet}$, $\tau_2/\tau_1$,
$C_2^\beta$, the jet charge, the planar flow, the number of jet
constituents, $\Delta R$ between subjets, sensitivity of trimmed
grooming, and the number of primary pp interaction vertices.
The MLP neural network is trained using a signal
sample from a SM Higgs-like resonance
decaying to a pair of longitudinally polarized W bosons
and a background sample of W+jets generated with \MADGRAPH,
splitting the events equally in training and test event samples to
compute the ROC curve.
The ROC curves obtained
from the multivariate methods are shown in Fig.~\ref{fig:rocMVA}.
Compared to the performance of $\tau_{2}/\tau_{1}$, a small
improvement is obtained using such multivariate discriminators.
This can be understood, because we find a large linear
correlation between $\tau_2/\tau_1$, which is the most sensitive
variable over a large range of efficiencies, and most of the other
observables.
We therefore focus in the following of this paper on a baseline tagger
based on $\tau_{2}/\tau_{1}$ and point out that, not considering
systematic uncertainties, there is potential gain in
using multivariate discriminators.

The comparison above is performed after requiring the pruned jet mass
to lie in the W boson mass window.  Since all substructure
variables are correlated with the jet mass, it is important to note
that the variable comparison shown in Fig.~\ref{fig:rocMVA} depends
strongly on the choice of the primary discriminant.  When the ungroomed
jet mass is the primary discriminant, a combination with other variables
provides a larger increase in discrimination, although the overall
performance is still inferior to the default choice of the pruned jet
mass and $\tau_2/\tau_1$.

\subsection{Performance in simulation}
\label{sec:efficiencies}

In this section we examine the simulated \pt and PU dependence of
the W tagging efficiency.  Efficiencies are defined for a pruned jet
mass of $60<m_{\text{jet}}<100$\GeV, and N-subjettiness
ratio of $\tau_2/\tau_1 < 0.5$.

\begin{figure}[htbp]
\centering
\includegraphics[width=0.48\textwidth]{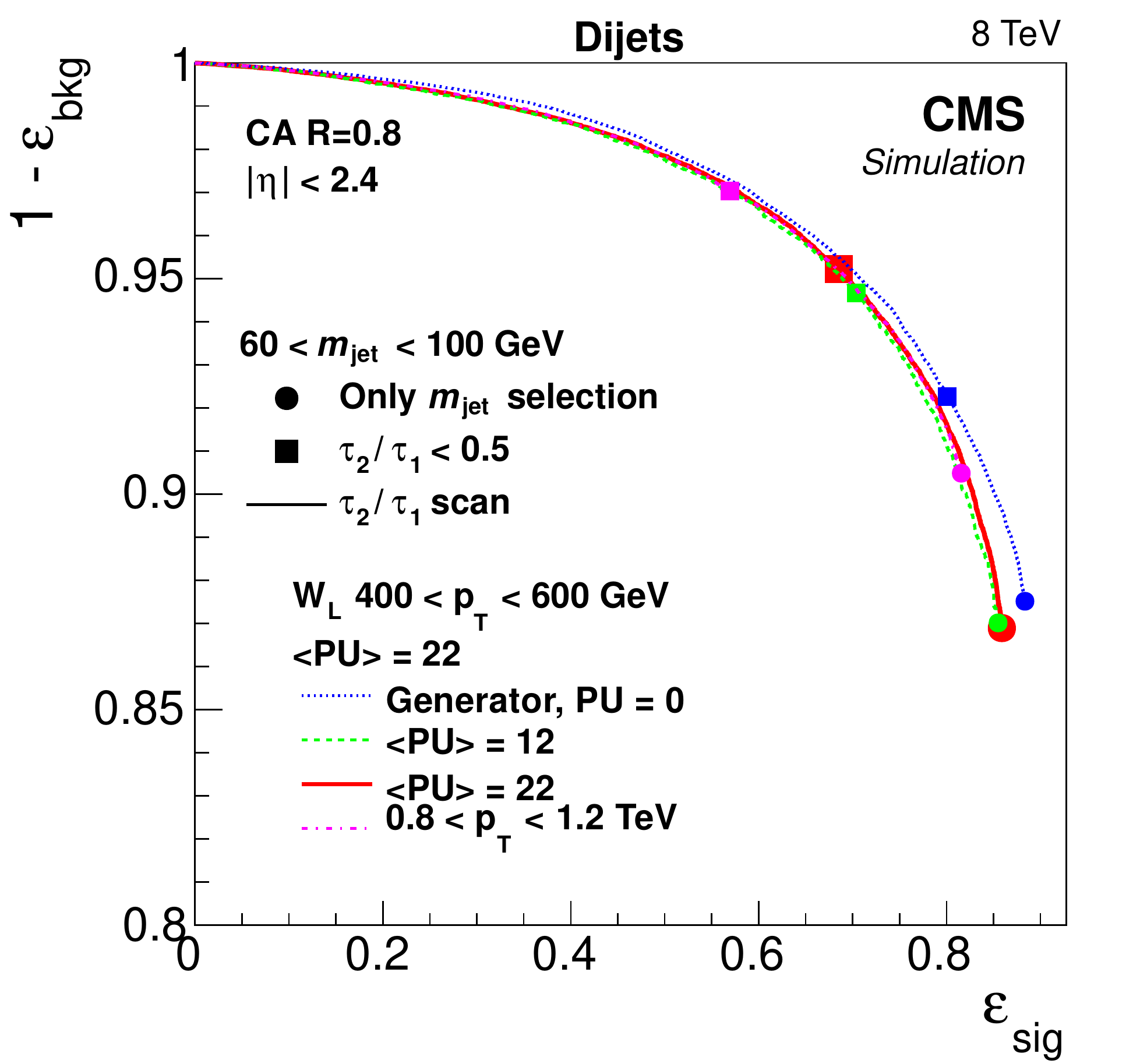}
\includegraphics[width=0.48\textwidth]{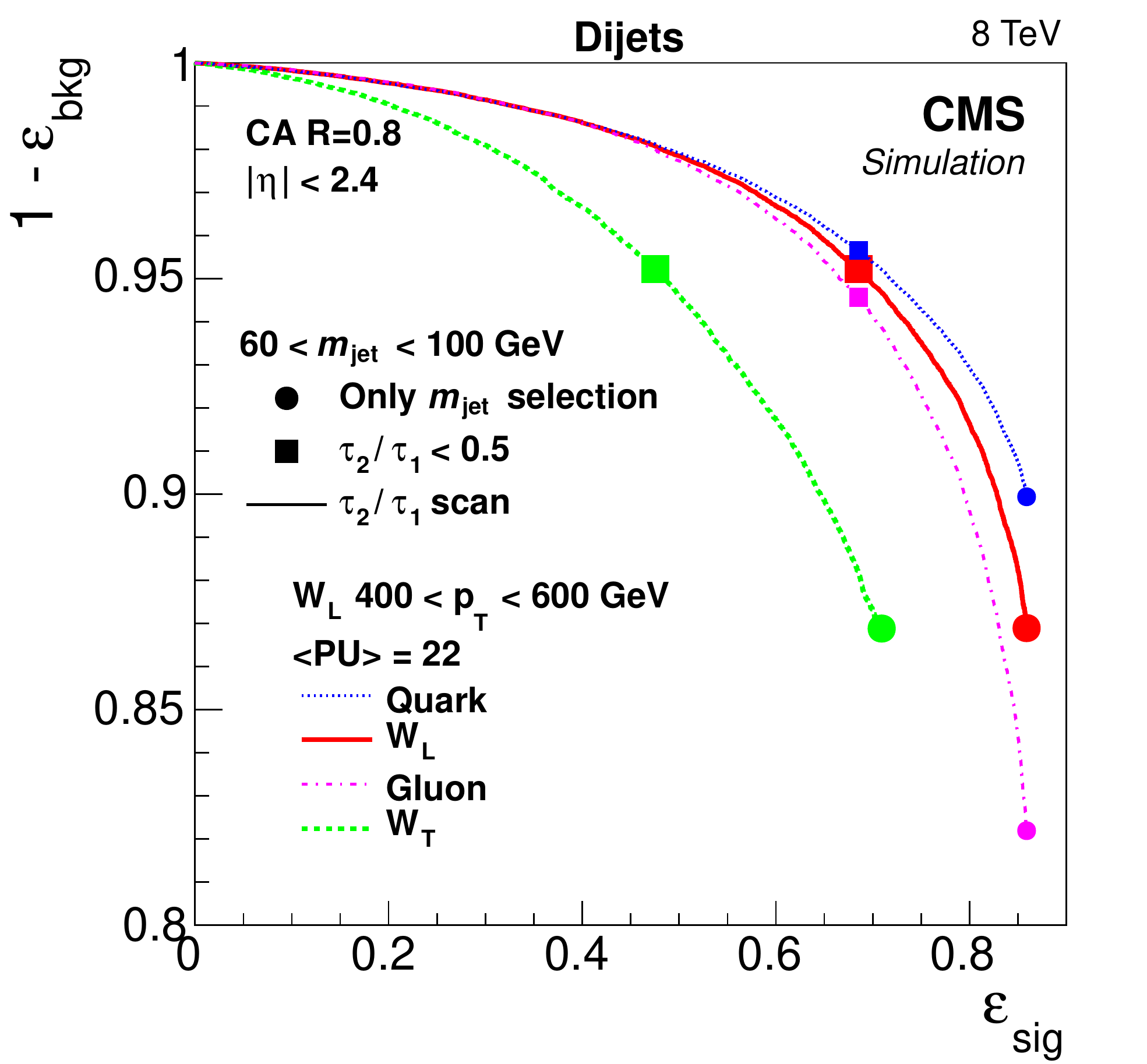}
\caption{Systematic effects on the performance of the pruned jet mass and $\tau_2 / \tau_1$ W-tagging algorithm
in the high jet \pt bin of 400--600\GeV.
The performance of the pruned jet mass selection $60 < m_{\text{jet}} < 100\GeV$ in the
various scenarios is indicated as a filled circle.
The performance of the combination of $60 < m_{\text{jet}} < 100\GeV$ and $\tau_2/\tau_1 < 0.5$
is indicated as a filled rectangle.
The lines correspond to the ROC curve of a selection on $\tau_2/\tau_1$ in addition to
$60 < m_{\text{jet}} < 100\GeV$.
The solid line corresponds (in both parts) to the standard scenario with an average of 22 pileup interactions
and longitudinally polarized W bosons ($\PW_\mathrm{L}$).
}
\label{fig:rocMVA2}
\end{figure}

In Fig.~\ref{fig:rocMVA2}, we compare systematic effects in terms of
change in the ROC response in the dijet final state for $400 < \pt < 600\GeV$.  In contrast to Fig.~\ref{fig:rocMVA}, where just
the performance of other variables was studied relative to that of
$m_{\text{jet}}$, here the efficiency is measured for the joint
condition on $m_{\text{jet}}$ and $\tau_2/\tau_1$, demonstrating the
impact of these discriminants.  The performance for the working point
requirements $60 < m_{\text{jet}} < 100$\GeV and
$\tau_2/\tau_1<0.5$ is also indicated.  The \HERWIG{++} sample is
used to model QCD jets, since we observe that it models the
pruned jet mass in data better than \PYTHIA~6 does.
Each of the displayed systematic effects is discussed below.

\begin{figure}[htbp]
\centering
\includegraphics[width=0.50\textwidth]{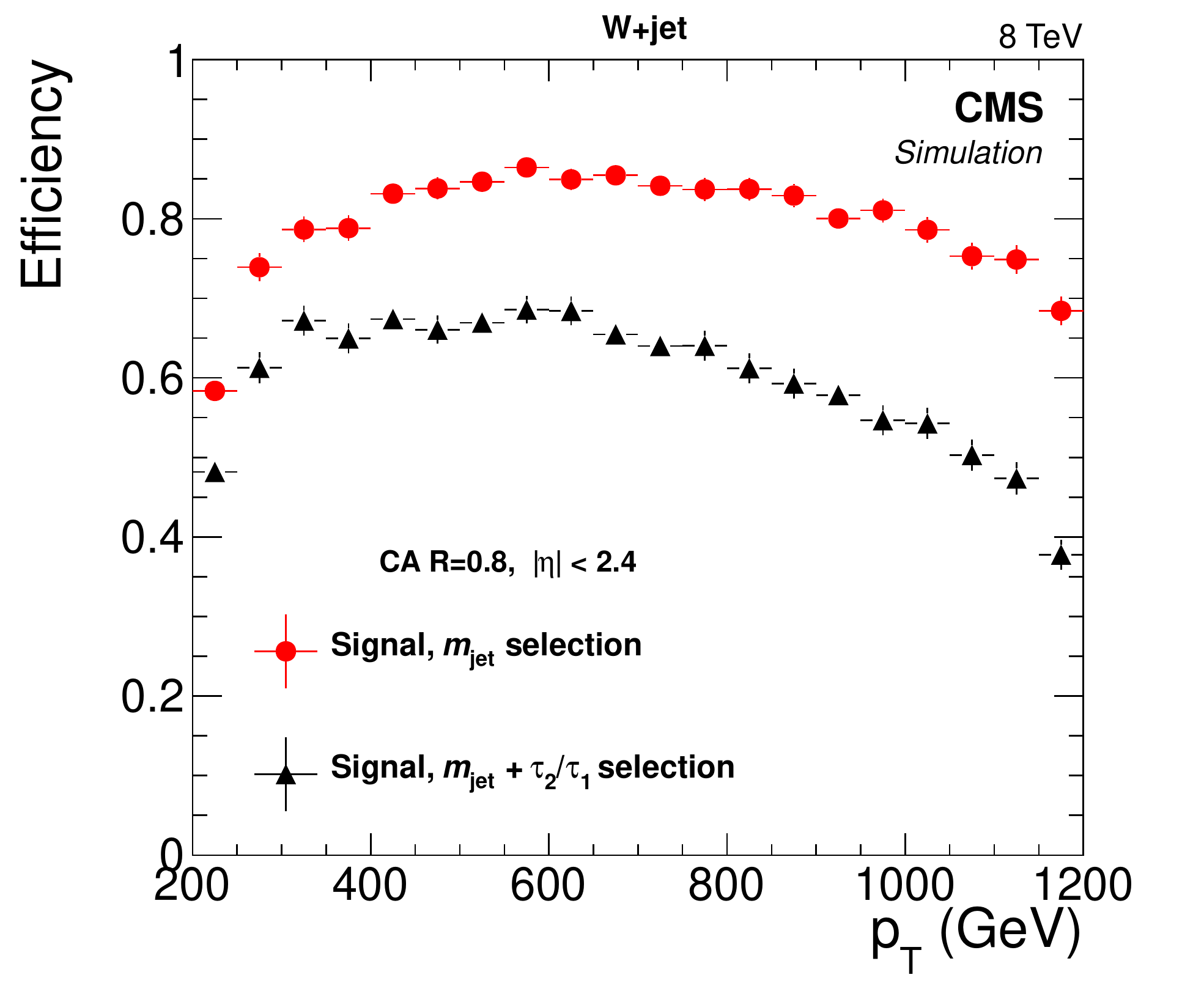}
\includegraphics[width=0.47\textwidth]{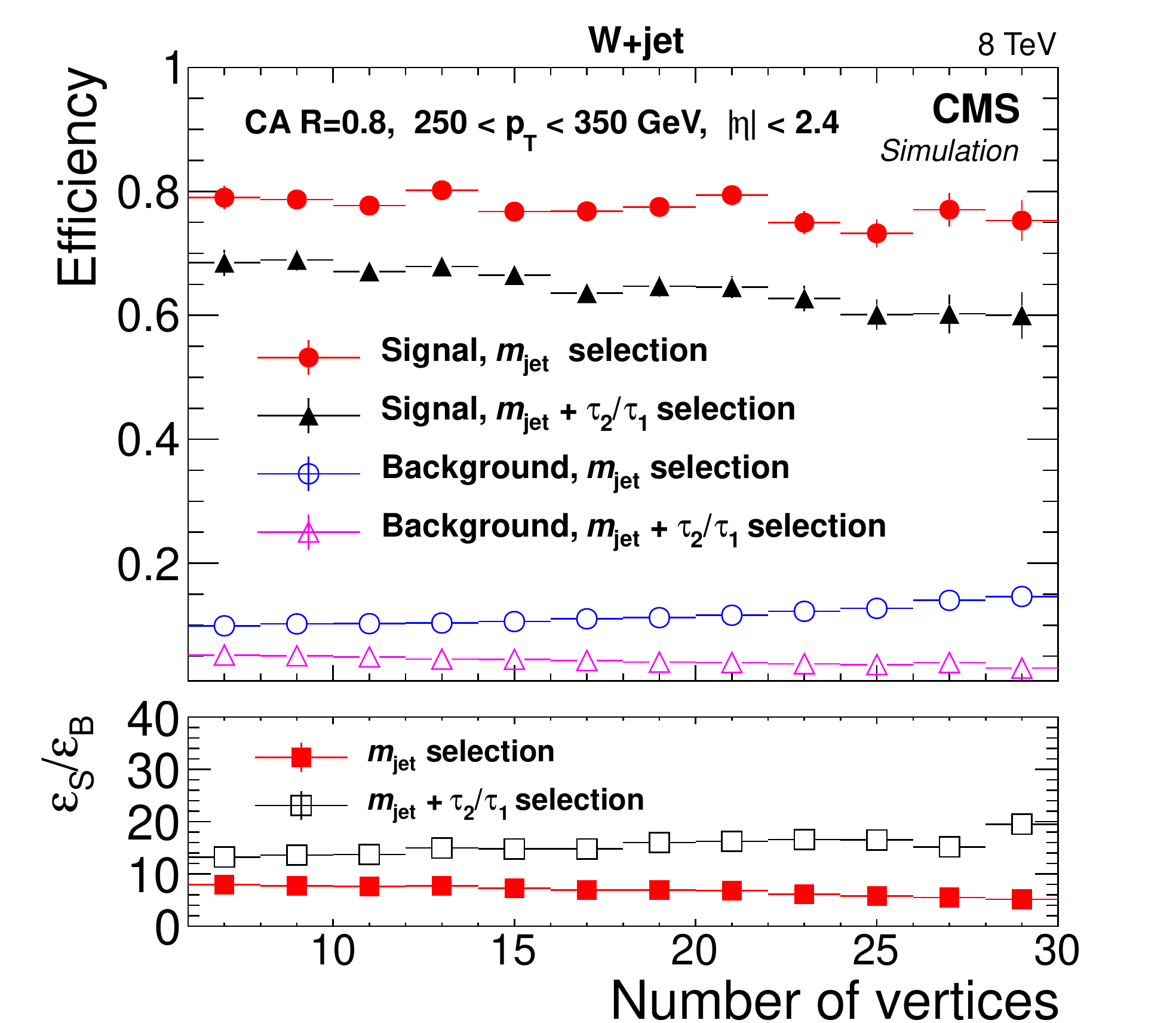}
\caption{Efficiency of the $m_{\text{jet}}$ selection and the combined $\tau_2/\tau_1$ and $m_{\text{jet}}$
selection on WW signal samples as a function of (left) \pt and (right) the number of reconstructed vertices.
The figure on the right also shows the mistagging rate for QCD jets estimated from the W+jets background sample.
The error bars represent the statistical uncertainty in the MC simulation and the horizontal ones the binning.}
\label{fig:efficiencyDijets}
\end{figure}

Figure~\ref{fig:efficiencyDijets} shows the efficiency of the baseline
selection ($60 < m_{\text{jet}} < 100$\GeV and
$\tau_2/\tau_1<0.5$) determined from a WW simulation.  The
efficiency is given as a function of (left) jet \pt and (right) the number of
reconstructed vertices, reflecting the contribution from pileup.
At low \pt, the efficiency increases with \pt for the same reason as
in Fig.~\ref{fig:ca8ak5}, namely that at higher \pt the showers from
the W decay quarks are more likely to be reconstructed within a
single CA8 jet.  Above 600\GeV, the efficiency begins to decrease as a
function of jet \pt, since at larger \pt the
PF candidate reconstruction degrades in resolving
the jet substructure and the pruning algorithm
therefore removes too large a fraction of the jet mass.
For Run II of the LHC, the particle flow reconstruction has been
optimized by making better usage of the
segmentation of the ECAL, where we expect
to maintain constant efficiency up to at least $\pt=3.5$\TeV~\cite{JME-14-002}.

The efficiency of the additional $\tau_2/\tau_1$ selection also drops as a function of \pt.
It is important to note that the same efficiency at an equivalent
background rejection rate can be reached
by adjusting the maximum $\tau_2/\tau_1$ as a function of \pt.
Figure~\ref{fig:rocMVA2}\,(left) shows that the ROC curve for jets with \pt
between 0.8 and 1.2\TeV (using a 2\TeV mass for the WW resonance)
is almost indistinguishable from the ROC curve derived from the
400--600\GeV \pt range, except that the working point corresponding to
$\tau_2/\tau_1<0.5$ (square) is at a lower signal efficiency.
Consequently, a fixed working point will degrade the efficiency with
increasing \pt.  However, by shifting the working point, the same
performance can be achieved.

The efficiency of the $m_{\text{jet}}$ selection as a
function of the number of reconstructed vertices, shown in
Fig.~\ref{fig:efficiencyDijets}\,(right), decreases by 6\% between 5 and 30
reconstructed vertices, whereas the additional $\tau_2/\tau_1$ selection efficiency
drops by 12\% over the same range.  However, the mistagging of the
background also decreases with pileup for the same selection, yielding
similar discrimination.
Efficiency and mistagging rate are affected by pileup in the same way,
since additional pileup shifts the $\tau_2/\tau_1$ distribution
towards higher values (towards background like) for both
signal and background.
Therefore, the same signal efficiency can
be reached at the same background rejection rate for up to 30
reconstructed vertices by merely adjusting the $\tau_2/\tau_1$ selection, as
demonstrated in Fig.~\ref{fig:rocMVA2}\,(left).  Moving from an average
pileup of 12 to 22 interactions shows almost no change in
the ROC response.

We also study the performance of jet substructure tagging algorithms
by convolving pileup, CMS detector resolution, and efficiencies in
reconstructing the particles that form the jets.  In
Fig.~\ref{fig:rocMVA2}\,(left), the generator level predictions without
pileup are compared with the performance after full CMS simulation
with pileup.  A small degradation is observed relative to generator
level, but the performance at detector level is almost as good as predicted at
particle level, although the W jet and the QCD jet
$\tau_2/\tau_1$ distributions are shifted up significantly by pileup
and detector effects, as seen in Fig.~\ref{fig:algorithms_jetmass}.

\subsection{W-polarization and quark-gluon composition}

\begin{figure}[th!b]
\centering
\includegraphics[width=0.50\textwidth]{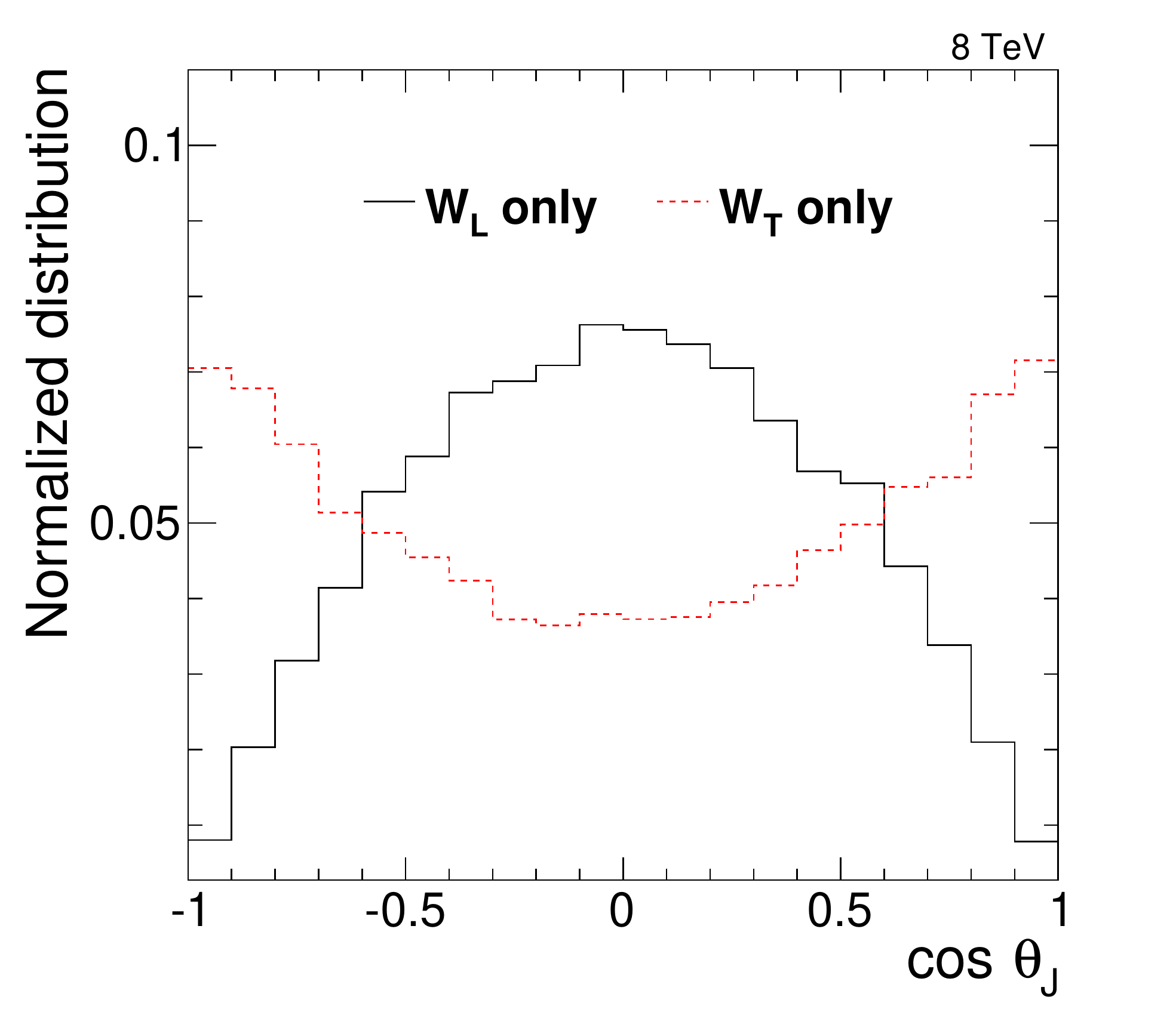}
\includegraphics[width=0.45\textwidth]{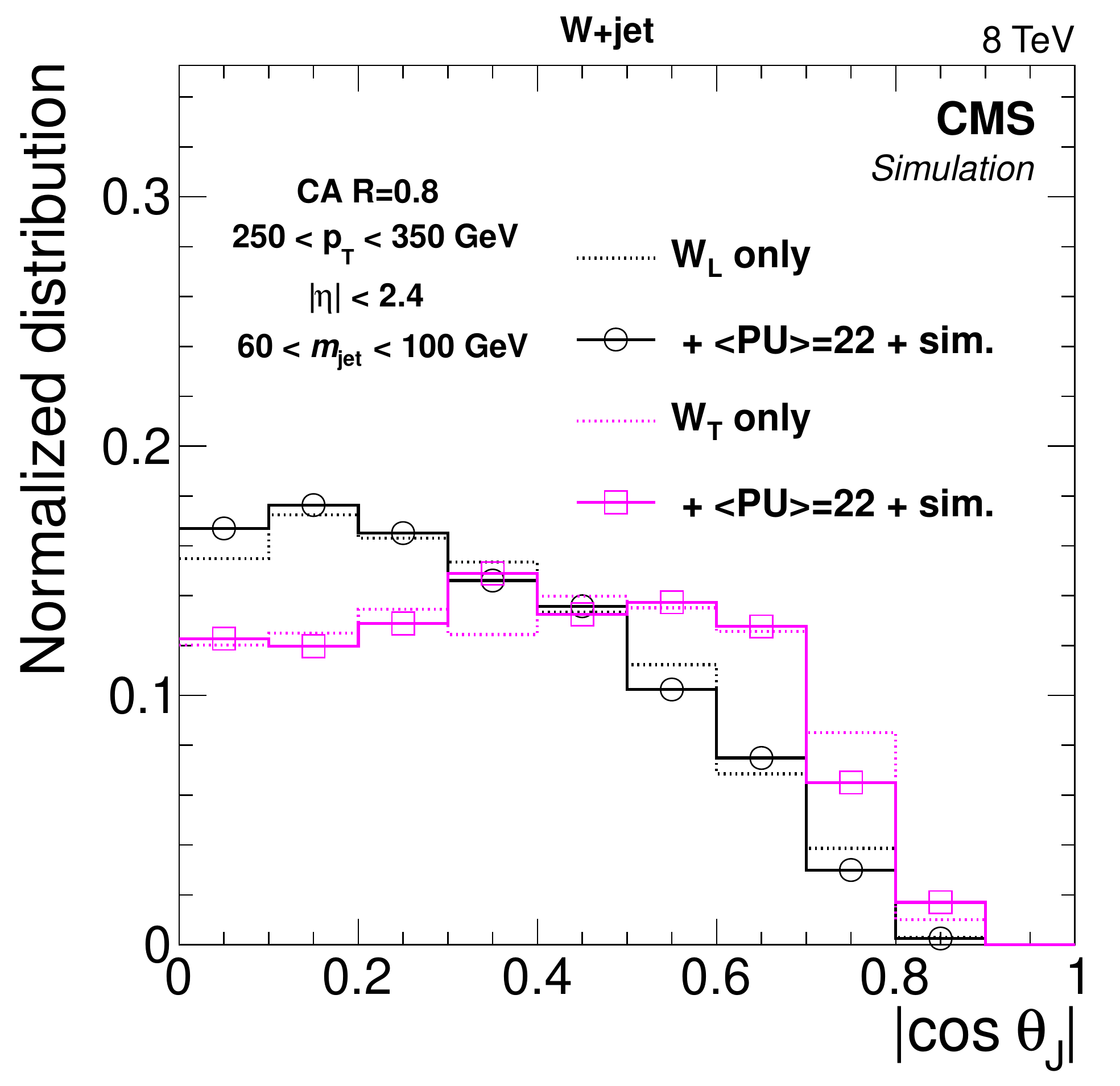}
\caption{(left) Generator level $\cos \theta_{J}$ distributions for longitudinally and transversely polarized W bosons.
(right) Subjet angular observables after a selection on pruned jet mass of $\PW_\mathrm{L}$ and $\PW_\mathrm{T}$ samples for jets with $250<\pt<350$\GeV.}
\label{fig:Wpol_GEN}
\end{figure}

An important factor that influences the W-tagging performance is the
polarization of the reconstructed W bosons.  Furthermore, the W
polarization can be used to identify the nature of any new phenomena,
such as, for example, through studies of new WW resonances, W boson helicities
at large \ttbar masses, or WW scattering.  We study the effect of W
polarization by comparing simulated samples of $X \to \PW\PW$, where the W
bosons are either purely longitudinally ($\PW_\mathrm{L}$) or transversely
($\PW_\mathrm{T}$) polarized.  The key observable is the helicity angle of $
\PW \to \cPaq\cPq'$ decays ($\cos \theta_{J}$) as defined in the rest frame of
the W boson relative to the W direction of
motion~\cite{Bolognesi:2012mm}.  The distribution of $\cos \theta_{J}$
at the parton level, where quarks are treated as final state
particles, is presented in Fig.~\ref{fig:Wpol_GEN}\,(left).  After
reconstruction, the polarization in W jets can be recovered using the
pruned subjets as a proxy for the W decay quarks.  However, using the
subjets, it is not possible to distinguish the fermion and
antifermion in the W decay, which restricts the distributions to
$0 \le \cos \theta_{J} \le 1$.
Figure~\ref{fig:Wpol_GEN}\,(right) shows the helicity angle between the two
pruned subjets for a 600\GeV X resonance, differing from
Fig.~\ref{fig:Wpol_GEN}\,(left) in that it includes reconstruction and
acceptance effects.
The depletion of events at $\abs{\cos \theta_{J}} \approx 1$ is due to two
acceptance effects. When $\theta_{J} \approx 0$, the partons would be
overlapping and thus reconstruction of two subjets is difficult.
When $\theta_{J} \approx \pi$, the one subjet tends to be much softer than
the other and this can cause the loss or misidentification of the subjet
originating from one of the W decay partons.
It appears that transversely polarized W bosons
decay with the quarks emitted closer to the direction of the W, and
therefore can be used to determine the polarization of the W boson.
Going further, the reconstructed $\cos \theta_{J}$ is compared to the
parton-level information.
The resolution on the angular distance between two subjets in the
laboratory frame is approximately 10\unit{mrad}, which translates to a
resolution of approximately 65\unit{mrad} on $\theta_{J}$ in the W rest
frame.  The resolution remains relatively constant over a large
range of W jet \pt.

Fig.~\ref{fig:rocMVA2}\,(right) compares the signal-to-background
discrimination of the W tagger for pure $\PW_\mathrm{L}$ and pure $\PW_\mathrm{T}$ signal
samples.  We observe that the pruned jet mass selection is less
efficient for $\PW_\mathrm{T}$; this is consistent with Fig.~\ref{fig:Wpol_GEN}\,(right), where the $\PW_\mathrm{T}$ jets with $|\cos\theta_{J}| \approx 1$ are removed
by the pruned jet mass selection.  This can be explained by a higher
asymmetry in the \pt of the two quarks from the $\PW_\mathrm{T}$ decay, such that
the pruning algorithm in a considerable fraction of events rejects the
particles from the lower \pt quark and yields a much lower jet mass.
In addition, the $\Delta R$ separation between the partons for pure
$\PW_\mathrm{L}$ bosons is smaller on average than for $\PW_\mathrm{T}$ bosons and is more
likely to be accepted by a CA8 jet.  Of the two effects, the dominant
contribution depends on the transverse momentum of the W jet.  For
higher jet \pt, the difference in the reconstructed $\cos \theta_{J}$
and $\Delta R$ between $\PW_\mathrm{L}$ and $\PW_\mathrm{T}$ becomes larger since the more
QCD-like topology of the transversely polarized W bosons becomes
important, i.e. it is easier to distinguish $\PW_\mathrm{L}$ and $\PW_\mathrm{T}$.  The
$\tau_2 / \tau_1$ discrimination power is also degraded for $\PW_\mathrm{T}$,
although, to a smaller degree than the pruned jet mass.

The composition of the QCD background also influences the
discrimination of the variables discussed in Section~\ref{sec:algo},
since the properties of quark- and gluon-initiated jets differ.
For example, gluon jets tend to have a larger jet mass than quark jets
and therefore fewer gluon jets are rejected by the pruned jet mass
selection; this can be seen in Fig.~\ref{fig:rocMVA2}\,(right).  On the contrary,
the $\tau_2/\tau_1$ discriminator rejects more gluon jets than quark
jets and for these reasons a similar performance for quarks and gluons
is achieved for the working point of $\tau_2/\tau_1<0.5$.

\section{Performance in data and systematic uncertainties}
\label{sec:efficiency}

\subsection{Comparison of data and simulation}
\label{sec:data}

We compare the distributions of substructure observables between simulation and data in inclusive dijet, W+jet
and $\ttbar$ samples.
The W+jet and dijet events are compared in respective jet \pt bins of 250--350\GeV and 400--600\GeV, and with
jets in the $\ttbar$ sample with $\pt>200\GeV$.
Simulation with different parton shower models of \PYTHIA~6, \PYTHIA~8 and \HERWIG{++} are also compared.

\begin{figure}[th!b]
\centering
\includegraphics[width=0.45\textwidth]{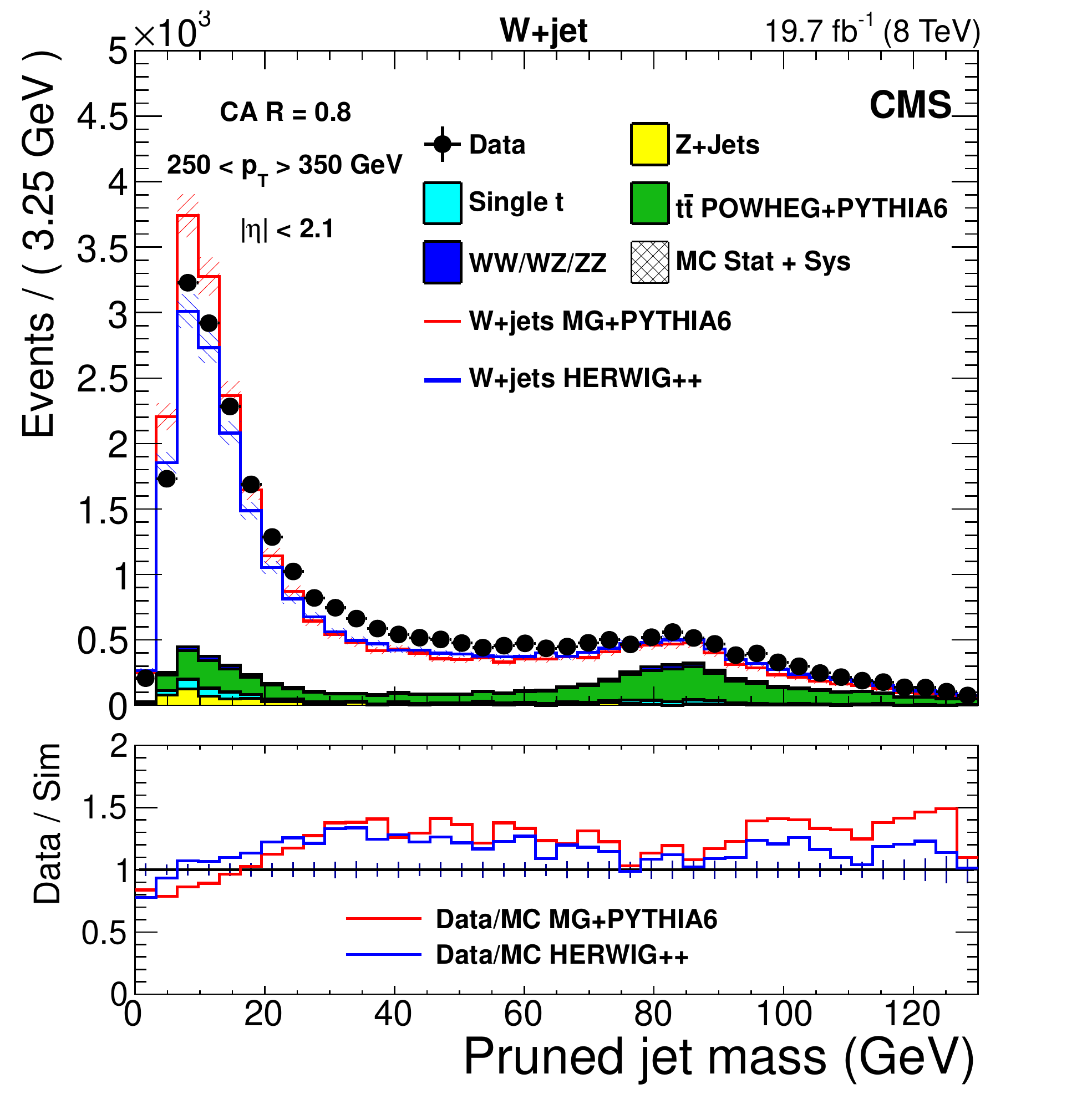}
\includegraphics[width=0.45\textwidth]{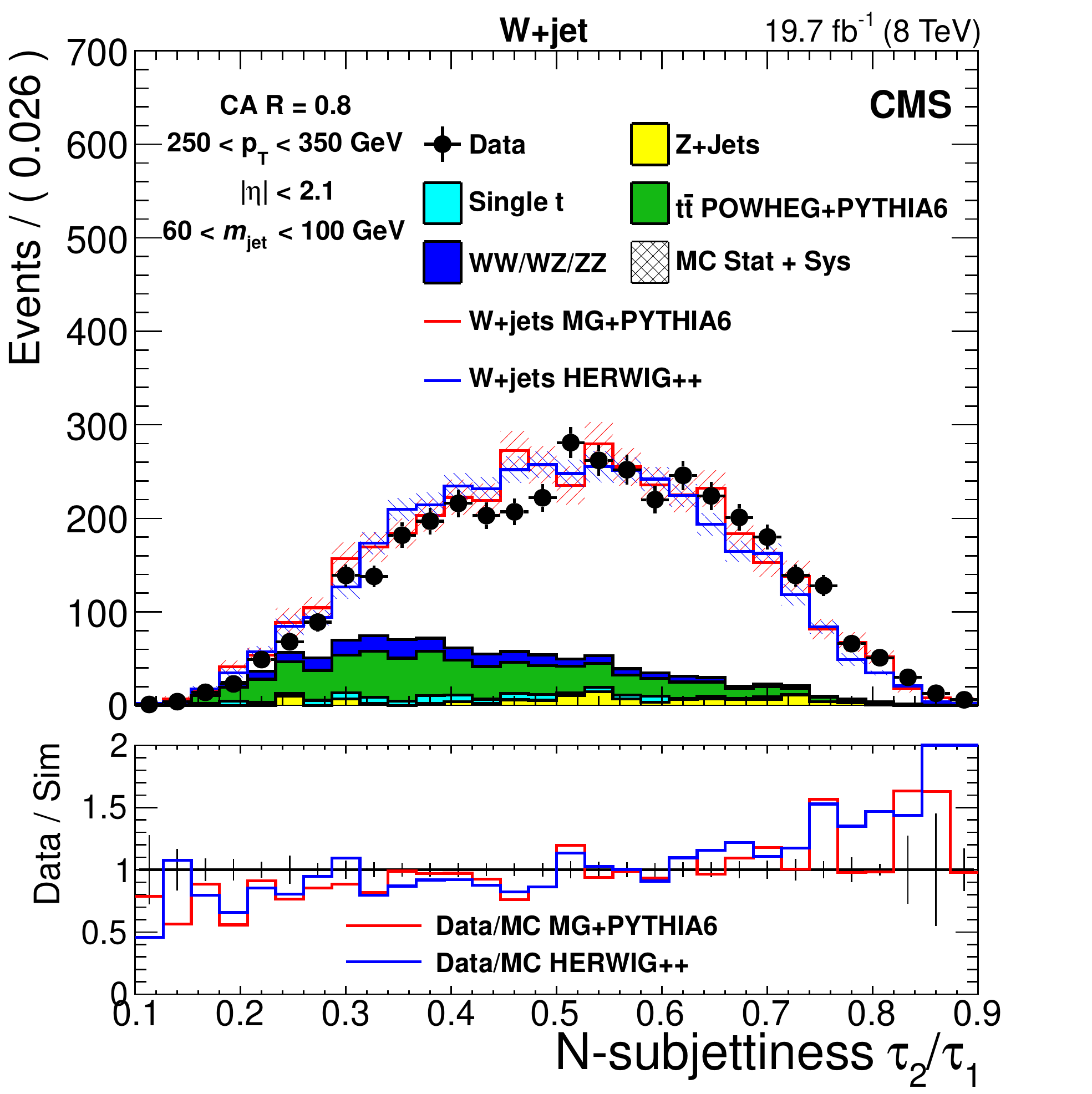}
\includegraphics[width=0.45\textwidth]{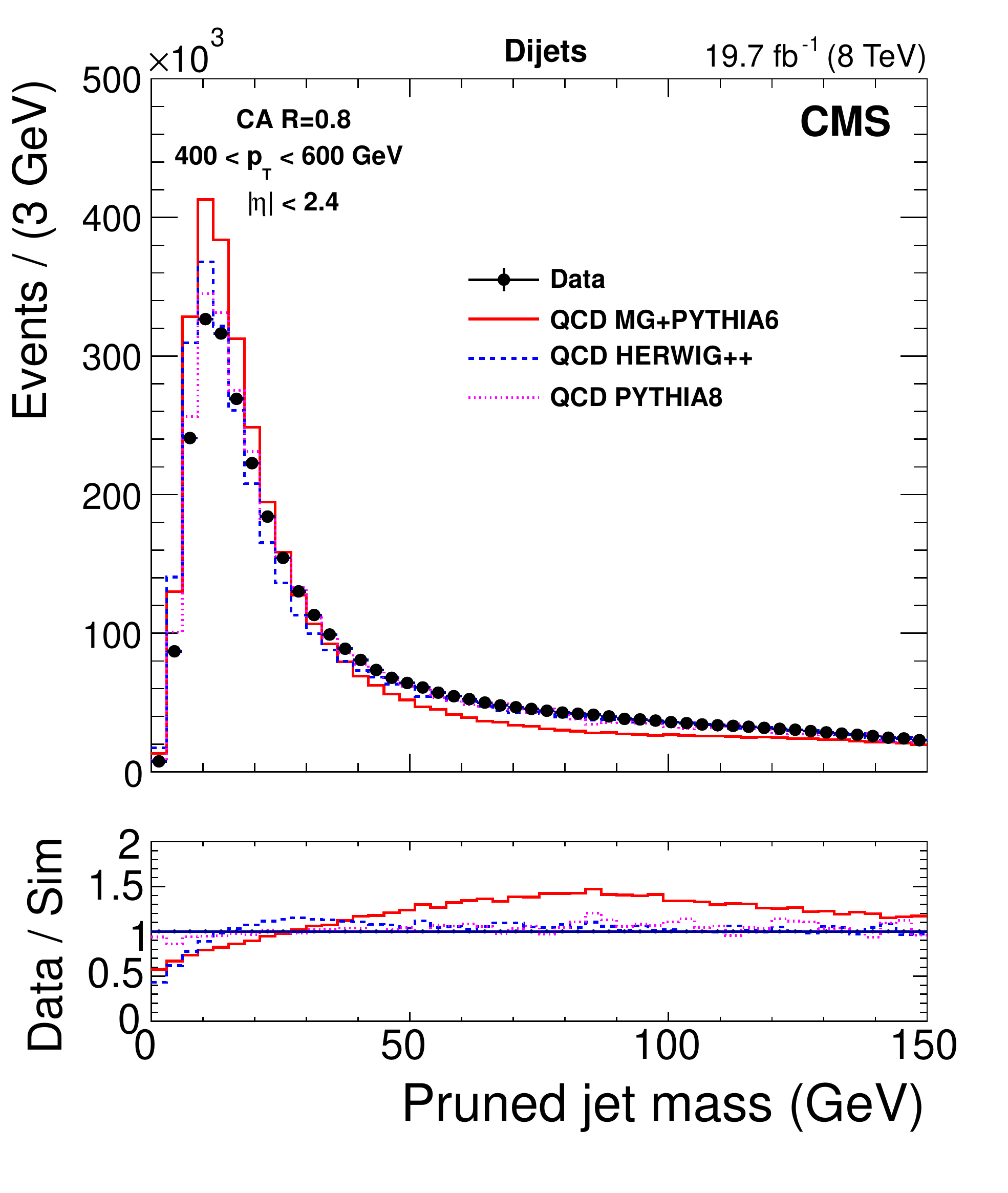}
\includegraphics[width=0.45\textwidth]{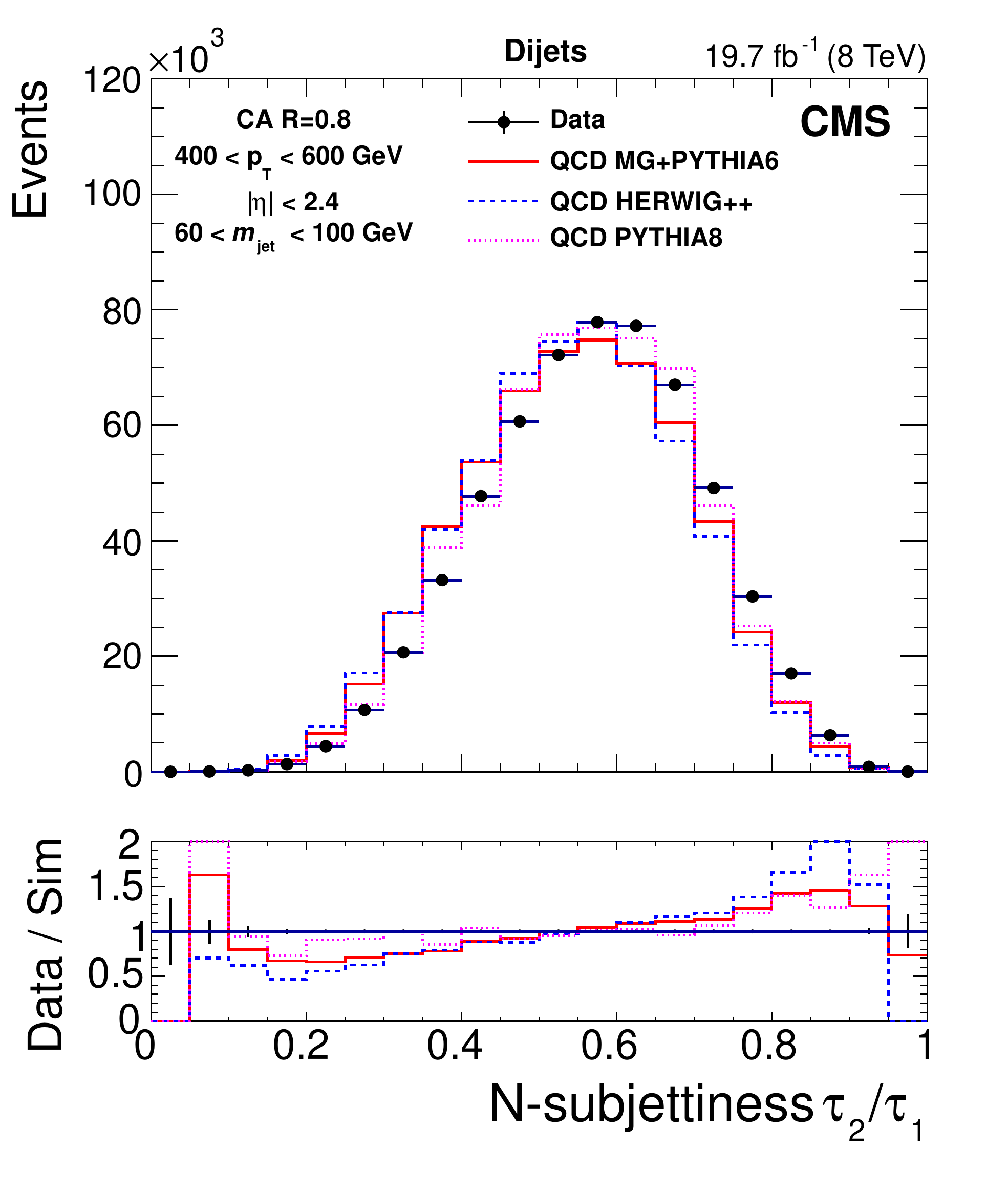} \caption{Pruned jet mass and N-subjettiness ratio $\tau_2 / \tau_1$ distributions in data and simulation for W+jets events in (upper left) and (upper right)
and for dijet events in (lower left) and (lower right). MG denotes the \MADGRAPH generator.
Below each figure the relative deviations are plotted between data and simulations.}
\label{fig:dj_datmc_mass}
\end{figure}

In Fig.~\ref{fig:dj_datmc_mass}, the pruned jet mass distribution is shown for both data and simulation in the dijet
and W+jet samples that probe the W-tagging variables using QCD jets.
We find that the agreement is good between data and simulation, but \HERWIG{++} agrees better than \PYTHIA~6,
and \PYTHIA~8 shows best agreement.
Similar findings have been reported in Ref.~\cite{topwtag_pas, jetmass, atlasjetmass}.
The $\tau_2 / \tau_1$ variable is also shown and found to agree better with \HERWIG{++} and best with \PYTHIA~8.

\begin{figure}[htbp]
\centering
\includegraphics[width=0.45\textwidth]{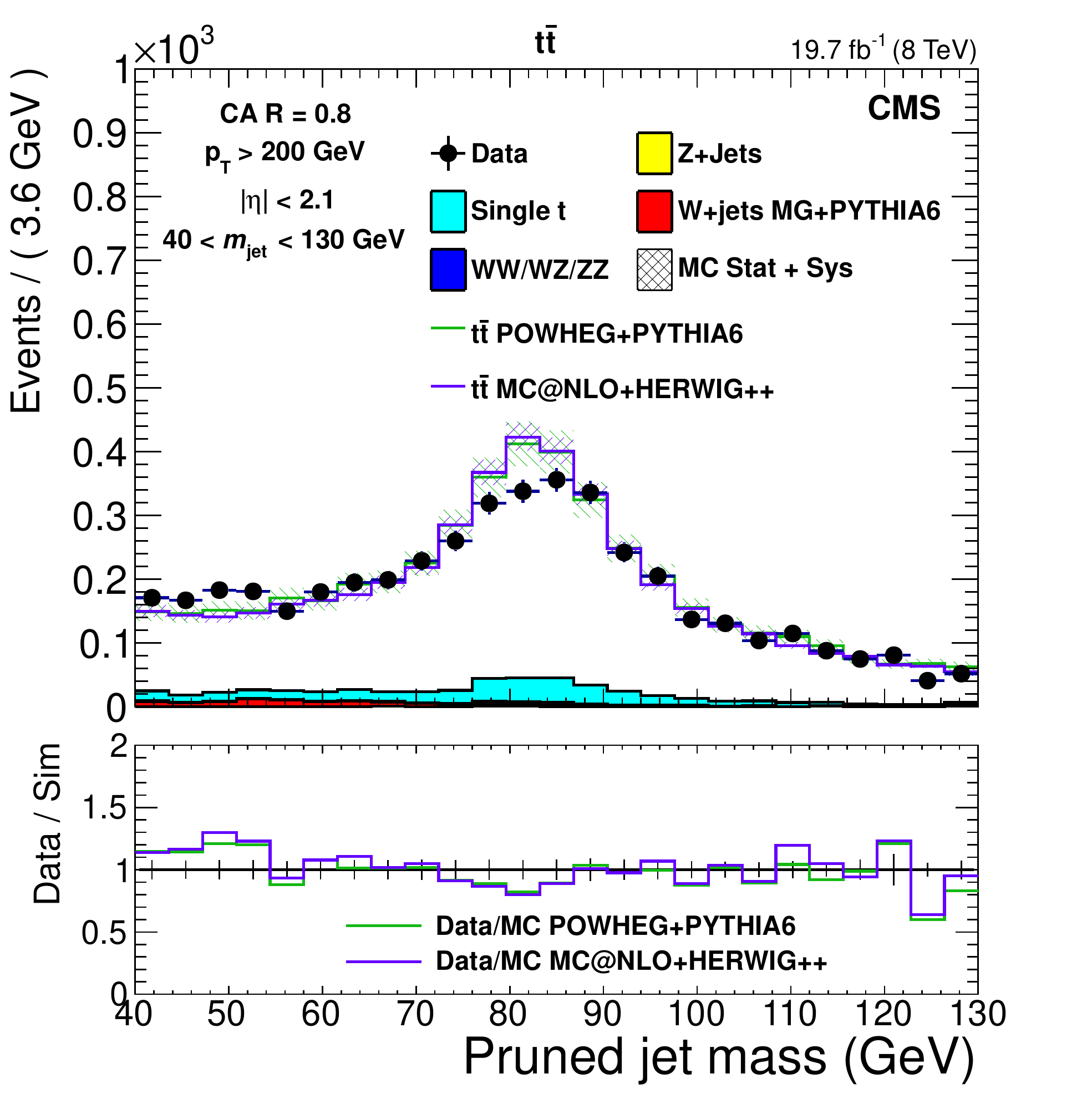} \includegraphics[width=0.45\textwidth]{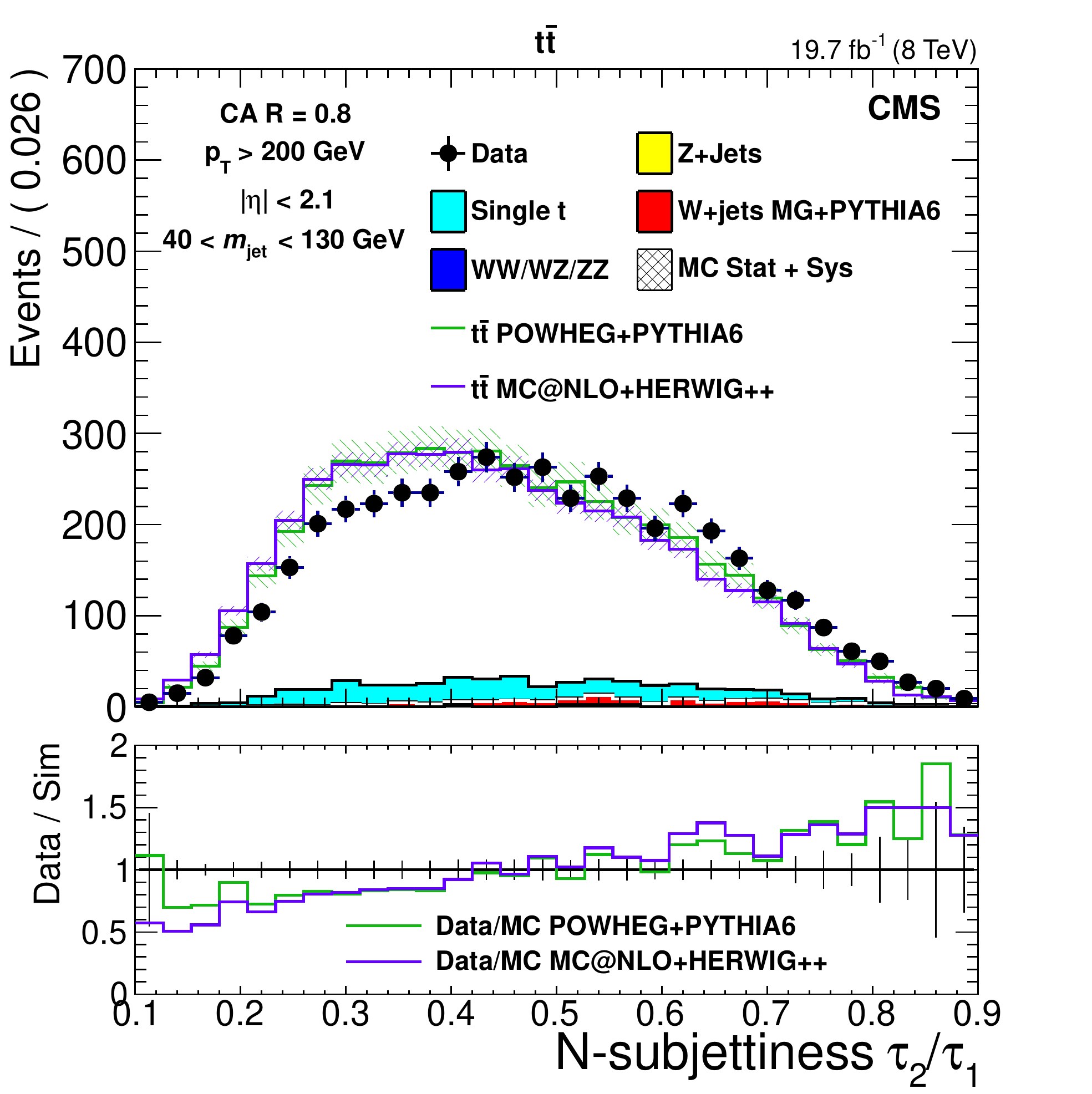}
\caption{Pruned jet mass and $\tau_2/\tau_1$ distributions for the lepton+jets $\ttbar$ control sample for the muon selection.
Below each figure the relative deviations are plotted between data and simulations.}
\label{fig:controlttbarkin_mu}
\end{figure}

To probe the description of W jets, we use the control sample of pure W bosons in the data from the high \pt lepton+jets $\ttbar$ sample.
The pruned jet mass and $\tau_2/\tau_1$ distributions in the $\ttbar$ control sample are shown in Fig.~\ref{fig:controlttbarkin_mu} for the muon selection.
The plots include systematic and statistical uncertainties, where the band of systematic uncertainty represents the normalization uncertainties on the VV, single top quark and W+jets
cross sections.
The systematic uncertainty is estimated to be 20\% determined from the relative difference in the mean value between the recent cross section measurement at $\sqrt{s}$ = 8\TeV at CMS
and the SM expectation~\cite{vvmeasurement}.
The agreement between simulation and data is reasonable, but there are discrepancies of the order of 10\%.
In Section~\ref{sec:scalefactors} we describe the derivation of data-to-simulation scale factors to correct for these discrepancies.
Generally, \POWHEG interfaced with \PYTHIA~6 provides a better description of the $\ttbar$ sample than \MCATNLO interfaced with \HERWIG{++}.

\begin{figure}[htbp]
\centering
\includegraphics[width=0.48\textwidth]{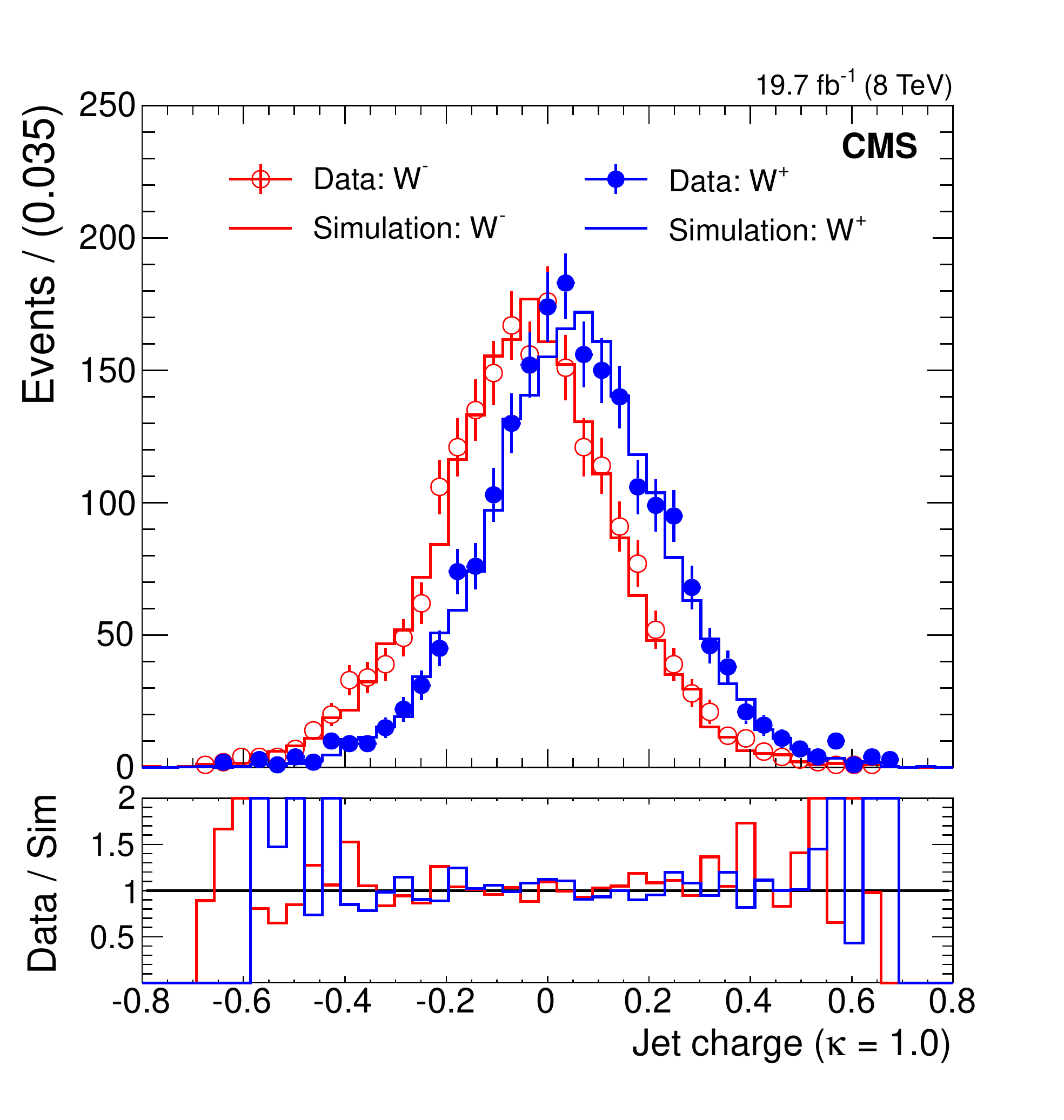}
\caption{Jet charge distributions in the $\ttbar$ control sample in simulation and data for $\PWp$ and $\PWm$ jets selected by requiring negatively and positively charge leptons, respectively.
Simulated distributions reflect the sum of $\ttbar$ (\POWHEG interfaced with \PYTHIA~6) and all other background processes.
Below each figure the relative deviations are plotted between data and simulations.}
\label{fig:ttbarControl_jetcharge}
\end{figure}

Finally, we compare the jet charge distribution of W jets in data and in simulation using the $\ttbar$ sample.
By selecting a negatively or positively charged lepton, we can effectively choose a $\PWp$ or $\PWm$ jet.
This can be seen in Fig.~\ref{fig:ttbarControl_jetcharge}.
While $\PWp$ and $\PWm$ jets can't be distinguished on an event-by-event basis,
their contributions to the $\ttbar$ data sample can be separated
with a significance larger than 5 standard deviations.
The jet charge distribution is well described by the simulation.

\subsection{Mistagging rate measurement}
\label{sec:fakerates}

\begin{figure}[htbp]
\centering
\includegraphics[width=0.45\textwidth]{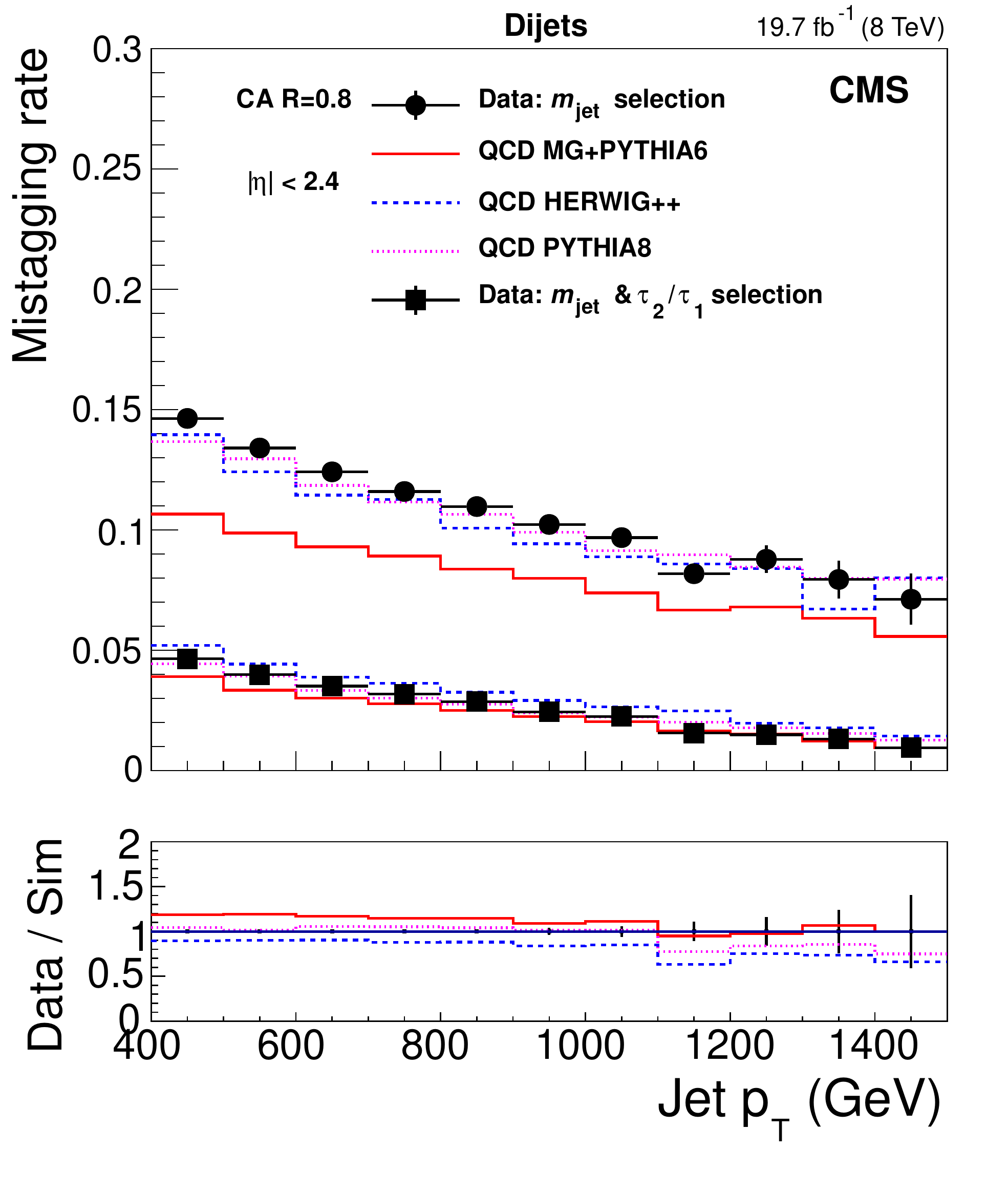} \includegraphics[width=0.45\textwidth]{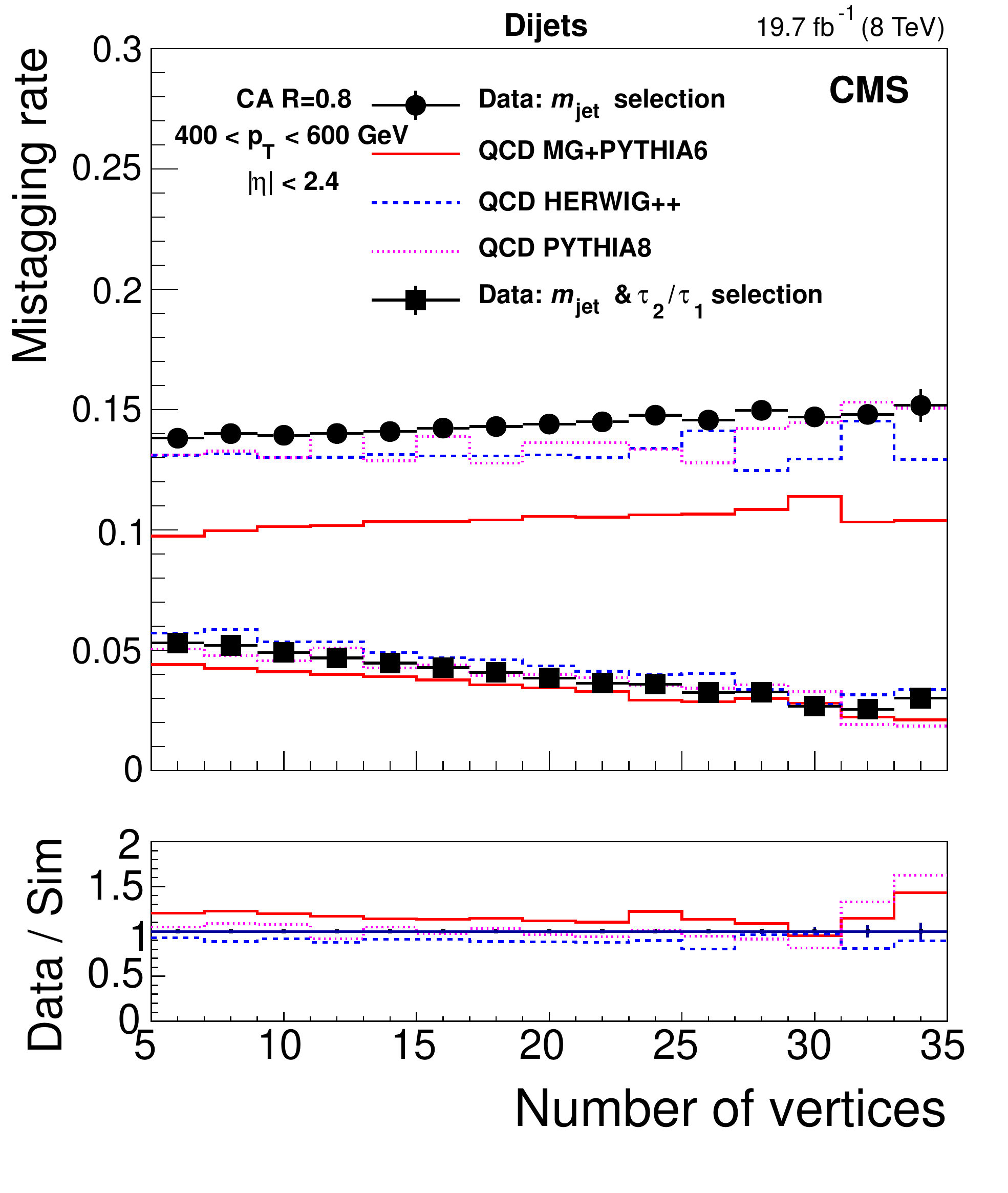}
\caption{Fraction of jets passing the $m_{\text{jet}}$ and $\tau_2/\tau_1$ selections in dijet data sample and
simulation as a function of (left) \pt and (right) the number of reconstructed vertices.
The data over simulation ratio is shown for the combination of the $m_{\text{jet}}$ and $\tau_2/\tau_1$ selections.}
\label{fig:fakerateDijets}
\end{figure}

A dijet sample is used to measure the rate of false positive W
tags, or mistags.  The mistagging rate is measured in data and
compared to simulation.  As discussed
previously, the W tagger selection requires
$60 < m_{\text{jet}} < 100$\GeV and $\tau_2/\tau_1<0.5$.
Figure~\ref{fig:fakerateDijets} shows the fraction of jets passing just the
$m_{\text{jet}}$ requirement, as well as
the simultaneous $m_{\text{jet}}$
and $\tau_2/\tau_1$ requirements, as a function
of \pt and of the number of reconstructed vertices.
Similarly as in the case of the
efficiency, the mistagging rate for the $m_{\text{jet}}$
and $\tau_2/\tau_1$ selections decreases as a function of \pt.
The mistagging rate of only the $m_{\text{jet}}$
requirement in data is well reproduced by \HERWIG{++} and
\PYTHIA~8, while \MADGRAPH{}+\PYTHIA~6 underestimates it.  When both
the $m_{\text{jet}}$ and $\tau_2/\tau_1$ requirements are
applied, the mistagging rate in data is reproduced better by
\PYTHIA~8 than by \MADGRAPH{}+\PYTHIA~6 and \HERWIG{++}.  The \pt
dependence in data is well reproduced by all generators.

As a function of pileup, the mistagging rate is stable within 1\% for
the $m_{\text{jet}}$ selection.  The mistagging rate for the
combination of the $m_{\text{jet}}$ and $\tau_2/\tau_1$
selections drops as a function of pileup as discussed in detail in Section~\ref{sec:efficiencies}.
The PU dependence is well reproduced by the simulation.

\subsection{Efficiency scale factors and mass scale/resolution measurement}
\label{sec:scalefactors}

The $\ttbar$ control sample is used to extract
data-to-simulation scale factors for the W jet efficiency.  These
factors are meant to correct the description of the W-tagging
efficiency in the simulation.  They depend on the definition of
the W-tagger as well as the MC generator used for simulation.  We
demonstrate the extraction of data-to-simulation scale factors for a
simple selection on $\tau_2 / \tau_1 < 0.5$, and jet mass scale, and
resolution based on a simulation using \POWHEG interfaced with \PYTHIA~6.
We are concerned only with the efficiency for the pure W jet
signal, and must therefore subtract background contributions to measure the
scale factors.
The pruned jet mass distribution is used to discriminate the pure W jet
signal from background contributions.
The generated W boson in the
$\ttbar$ simulation provides a model of the
contribution from the W jet peak in the pruned jet mass.
The contribution from combinatorial background is
derived from $\ttbar$ simulation as well.
This model is fitted directly in the distributions of data and in their simulation.

\begin{figure}[htbp]
\centering
\includegraphics[width=0.45\textwidth]{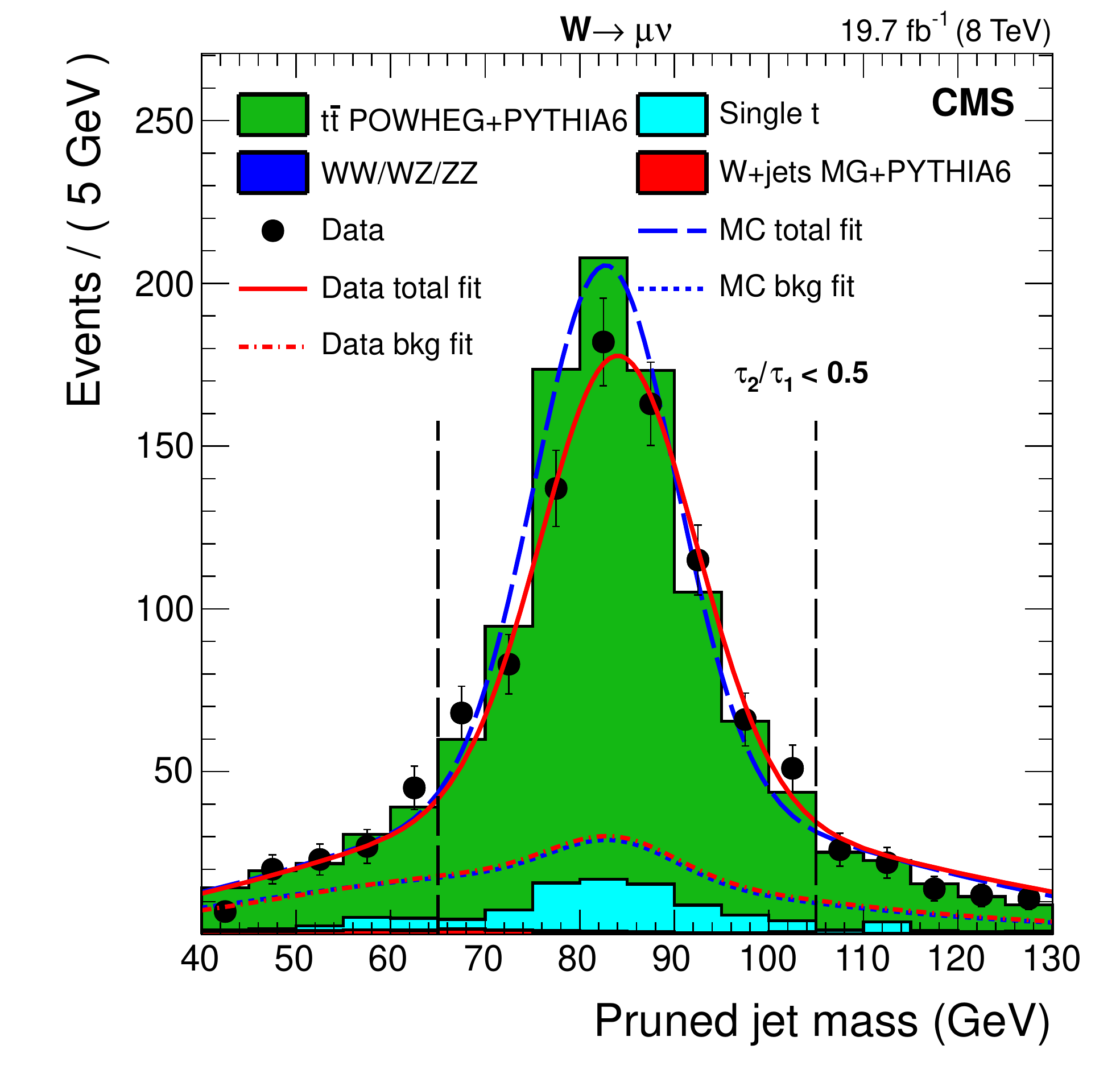}
\includegraphics[width=0.45\textwidth]{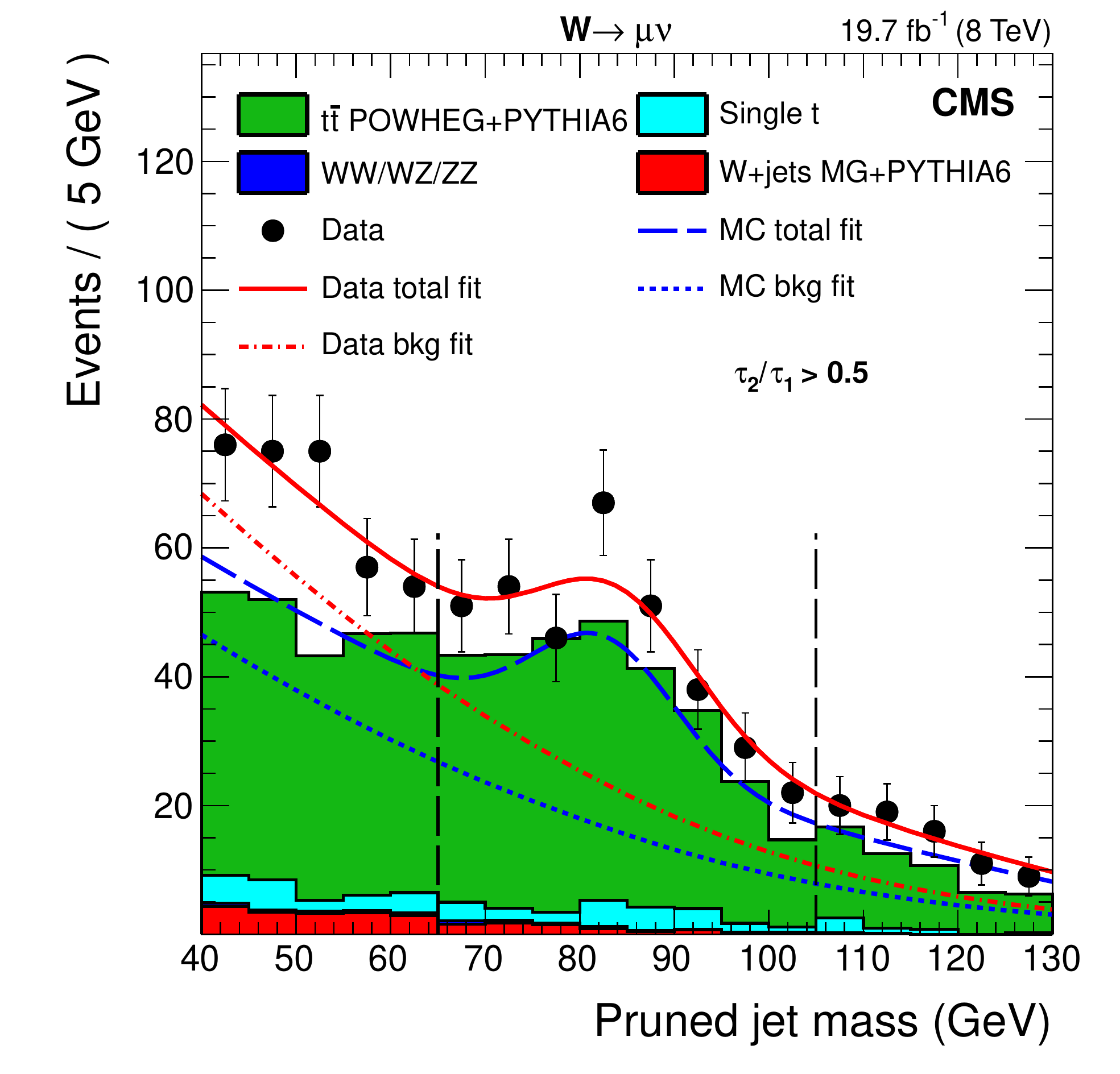}
\includegraphics[width=0.45\textwidth]{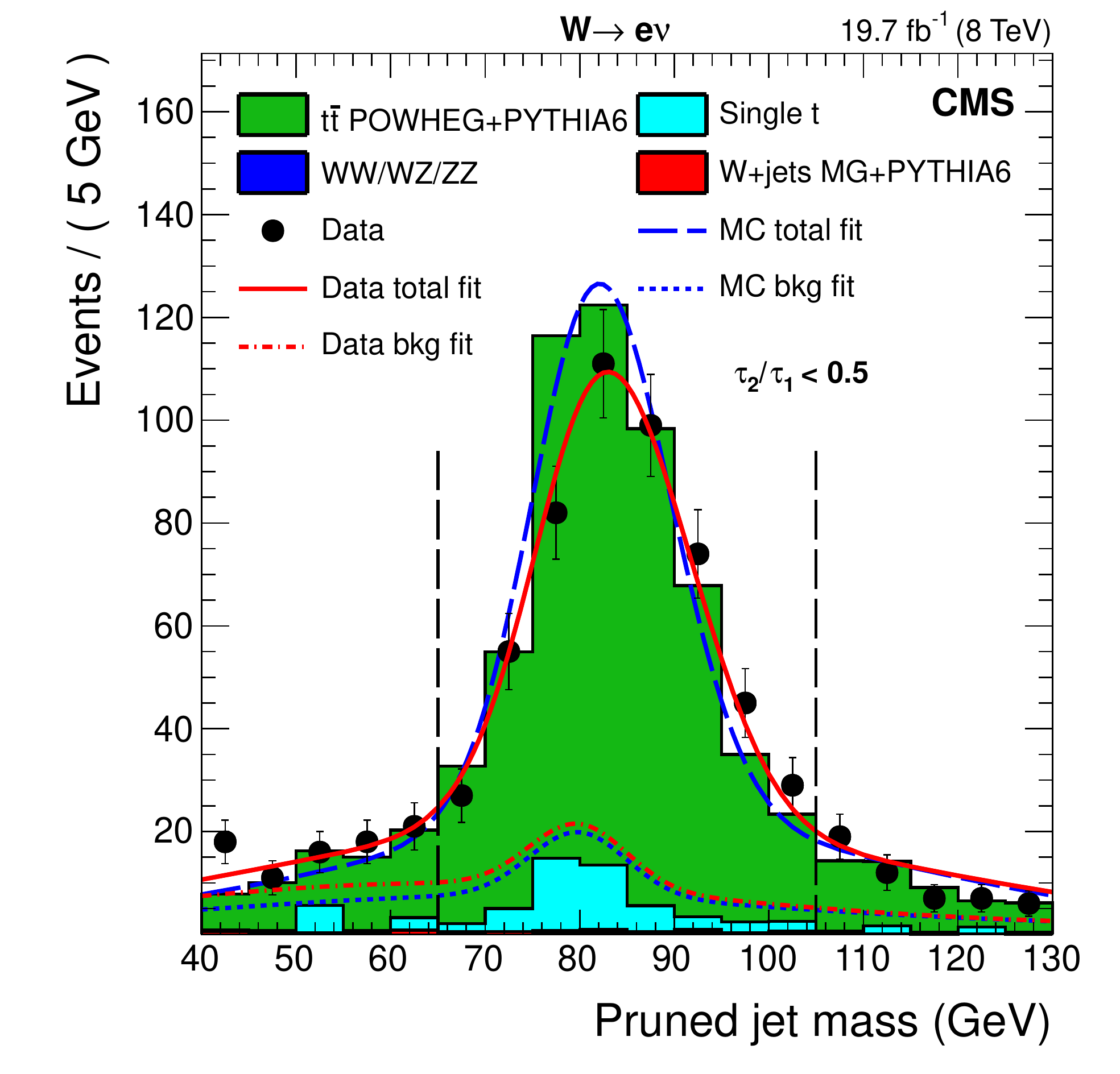} \includegraphics[width=0.45\textwidth]{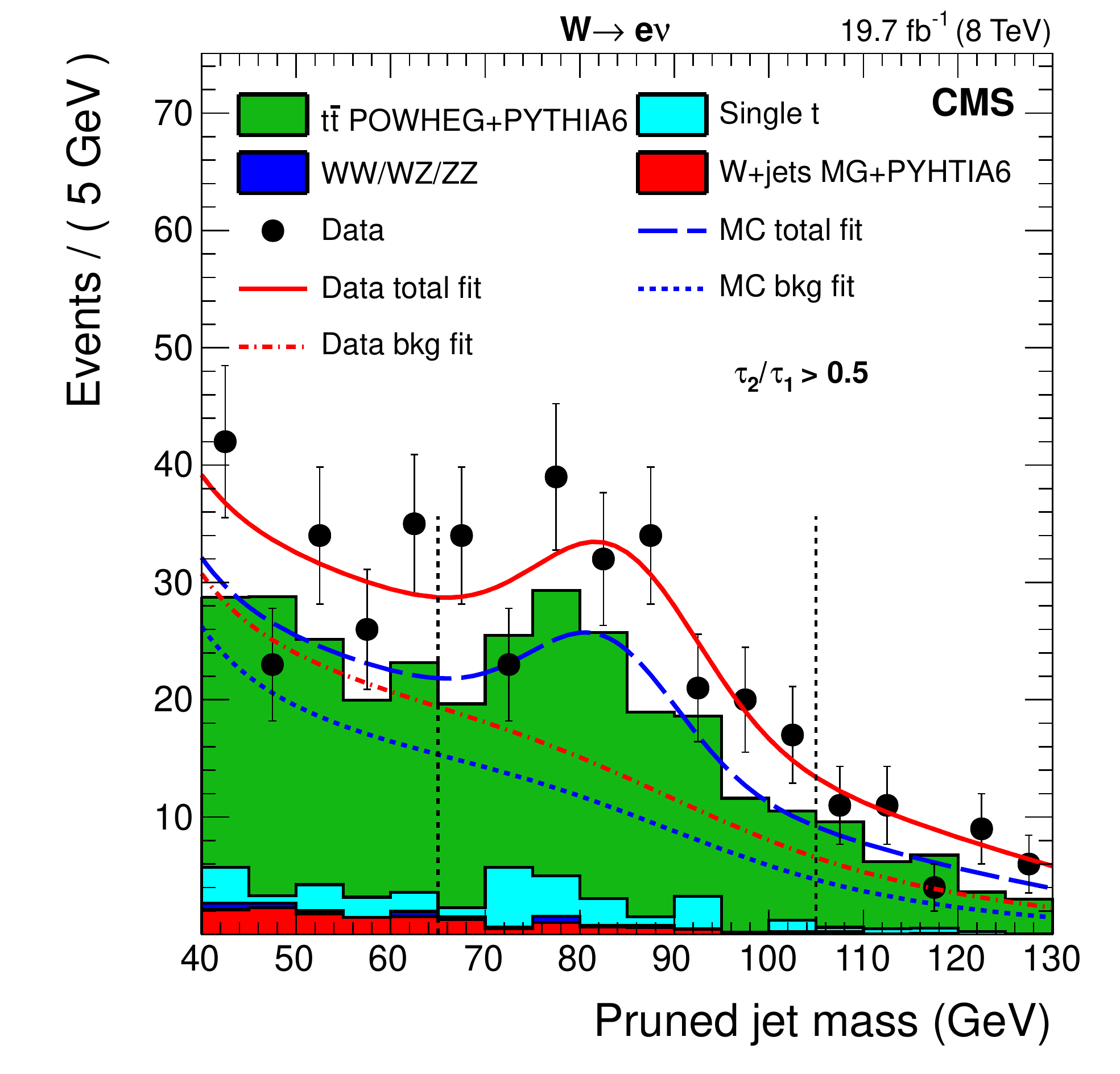}
\caption{Pruned jet mass distribution in the $\ttbar$ control sample that (left column) pass and (right column) fail the $\tau_2 / \tau_1 < 0.5$ selection for the (upper row) muon, and for the (lower row) electron channels.
The result of the fit to data and simulation are shown, respectively, by the solid and long-dashed line and the background components of the fit are shown as dashed-dotted and short-dashed line.}
\label{fig:ttbarControl_nocut}
\end{figure}

The scale factors (SF) for the selection on $\tau_2 / \tau_1 < 0.5$ are
extracted by estimating the selection efficiency on both data and
simulation.  The pruned jet mass distribution of events that pass and fail
the $\tau_2 / \tau_1$ selection are fitted
simultaneously to extract the selection efficiency on the pure W jet
component as shown in Fig.~\ref{fig:ttbarControl_nocut}.
The ratio of data and simulation efficiencies are taken as
the W-tagging efficiency SF.  In the $\ttbar$ control
region we use a mass window of 65--105\GeV,
because of a slight shift in the mean mass
of the W boson peak in $\ttbar$ events of $\approx$1.5\GeV.
In simulation the slight shift in mass is found to be
primarily due to extra radiation in the W jet from the nearby b quark.
Additional requirements to reduce the combinatorial background from
$\ttbar$ improve the precision of the determined scale
factor.  Therefore, the angular distance $\Delta R$ between the W jet candidate
and the closest b-tagged AK5 jet is required to be less than 2.0,
which is typical for highly boosted top quark decays~\cite{topwtag_pas}.
This additional selection reduces the uncertainty on the scale factor
by 21\%.  Further reduction of the combinatorial background can be
achieved through requirements on top quark masses, but
the limited number of $\ttbar$ events suggests that
this can become relevant only with a larger data sample.
The results of the fit are shown in Fig.~\ref{fig:ttbarControl_nocut}.
We find the ``pass" sample agrees well between the data and simulation
while the ``fail" sample is not as well modeled,
particularly when the failing jet is not a fully merged W boson
but a quark or gluon jet.
This is compensated in our computation of the data-to-MC scale factor.
The scale factor is computed to be $0.93 \pm 0.06$.  The uncertainty in
the SF is purely statistical.  In
Section~\ref{sec:syseff}, we discuss systematic effects to this scale
factor.
The \pt dependence of the scale factor was also studied at a limited
statistical precision. In two \pt bins between 200--265 and 265--600\GeV
the scale factors were found to be $1.00 \pm 0.09$ and $0.92 \pm 0.10$, respectively.
No significant \pt dependence of the scale factor is observed.

To extract corrections to the jet mass scale and resolution, we use the
mean $\langle m \rangle$ and resolution $\sigma$ value of the Gaussian
component of the fitted function of the W bosons in the passed
sample.  Since we do not expect the jet mass scale and resolution
to differ between electron and muon channels, the muon and
electron data are fitted simultaneously, forcing the $\langle m \rangle$
and $\sigma$ of the Gaussian component of the fit to be the same
in the two channels. The fits are shown for the
$\tau_2 / \tau_1 < 0.5$ selection in Fig.~\ref{fig:ttbarControl_nocut}\,(left column), and the
resulting parameters are summarized in Table~\ref{tab:params}.  We find
that both the W jet mass scale and resolution in data are larger than
that in simulation.  In the simulation $\langle m \rangle$ must
therefore be shifted by $1.7 \pm 0.6\%$ and $\sigma$ be
enlarged by $11 \pm 9\%$ to correct for the difference between data
and simulation.

\begin{table}[!htb]
\centering
\topcaption{Summary of the fitted W-mass peak fit parameters.}
\label{tab:params}
 \begin{tabular}{c|ccc}
  Parameter & Data & Simulation & Data/Simulation \\
  \hline
  $\langle m \rangle$ &$84.1 \pm 0.4\GeV$ & $82.7 \pm 0.3\GeV$ & $1.017 \pm 0.006$ \\
  $\sigma$ & \ $8.4 \pm 0.6\GeV$ & \ $7.6 \pm 0.4\GeV$ & $1.11 \pm 0.09$ \\
 \end{tabular}

\end{table}

\subsection{Systematic uncertainties}
\label{sec:syseff}

We now discuss systematic uncertainties in the W tagging scale factor.
Several important effects, including the modeling of parton shower and the PDF,
polarization of the W boson, the pileup, presence of nearby jets, the jet mass scale,
jet energy scale, and resolution effects, as well as less dominant contributions from
the uncertainties in lepton identification, b tagging and \MET scale
are considered.  The effects from the modeling of the parton shower and the PDF
are quantified by the difference between the efficiency in a \ttbar
sample generated with \POWHEG interfaced with \PYTHIA~6 and a
sample from \MCATNLO interfaced with \HERWIG{++}.  The effects
from modeling the underlying event in the simulation are estimated by
comparing three alternative tunes (Z2*, AMBT1~\cite{ambt1} and
AMBT2~\cite{ambt2}) of the multiple parton interaction model in the
\PYTHIA~6 simulation, and taking the maximal observed difference
as an estimate of the uncertainty.

As we have shown above, the polarization of the W boson has a
significant impact on the W-tagging efficiency, which has to be taken
into account when propagating the scale factor from
\ttbar events to other final states.  The W boson polarization in
\ttbar events has been measured by CMS with a
precision of 4\%~\cite{wpol}.  Although the agreement in this result
between data and theory was found to be less than the quoted
precision, we nevertheless use this number as an upper limit and
include it into an uncertainty of the W tagging scale factor.

The effect from jet mass scale and resolution is evaluated by changing
them by the uncertainty in the fitted mean and
resolution, estimated in Section~\ref{sec:scalefactors}.  Their impact on
the W-tagging efficiency is small, since the W boson mass peak is well
within the chosen pruned jet mass window.

The jet energy scale and resolution are changed within
their \pt- and $\eta$-dependent
uncertainties~\cite{Collaboration:2012dp,JME-JINST}.  The impact of
nearby jets on the scale factor is estimated by comparing it to a scale
factor for an explicit requirement on the angular distance between the
closest AK5 jet and the W jet of $\Delta R > 1.3$.  The uncertainty
from pileup is determined by moving the minimum bias cross section
within its measured uncertainty of 6\%~\cite{minbiasxsec}.
The scale factors for lepton and b jet
identification are also changed within their uncertainties.  Finally,
uncertainties in the energy and momentum scale and resolution of
leptons and jets in the event are propagated to an uncertainty on the \MET.

The results are summarized in Table~\ref{tab:systematics}.  The
dominant systematic effect on the scale factor for the efficiency is
from modeling of the parton shower and PDF, with a systematic uncertainty
of 6.0\%.  The quadratic sum of systematic uncertainties of 7.6\% is
comparable to the statistical uncertainty on the scale factor of
6.4\%.

\begin{table}[!htb]
\centering
\topcaption{Summary of uncertainties on the W jet identification efficiency scale factor.}
\label{tab:systematics}
 \begin{tabular}{c|c}
  Source & Effect on the scale factor \\
  \hline
  Parton showering + PDF & 6.0\% \\
  Underlying event &  $<$0.5\% \\
  W-polarization & 2.0\% \\
  Pileup & 1.8\% \\
  Nearby jets & 2.4\% \\
  Jet mass scale &  $<$0.5\% \\
  Jet mass resolution & 1.9\% \\
  Jet energy scale &  1.9\% \\
  Jet energy resolution & 0.9\% \\
  Lepton ID &  $<$0.5\% \\
  b-tagging &  $<$0.5\% \\
  \MET &  $<$0.5\% \\
  \hline
  Total systematic & 7.6\% \\
  \hline
  Statistical & 6.4\% \\
  \hline
  Total & 9.9\% \\
 \end{tabular}

\end{table}

\section{Summary and outlook}
\label{sec:summary}

In this paper we presented techniques for the identification of jets originating
from highly boosted W bosons that decay into $\cPaq\cPq'$, where
the final decay products are reconstructed within a single jet, called
a W jet.  The pruned jet mass, used as the primary identifying
observable for W jets, and several substructure observables that can
provide additional signal to background discrimination, were evaluated
for their impact.

The investigated substructure observables were three variants of the N-subjettiness $\tau_2 / \tau_1$, the mass drop, the Qjet volatility, the double ratio of the energy-correlation function $C_{2}^\beta$, and the jet charge.
Effects from pileup, detector resolution, polarization of the W boson, and the quark/gluon composition of QCD jets, as well as the performance of the discriminant at large \pt were studied.
The results were evaluated after applying a pruned jet mass selection, and the $\tau_2 / \tau_1$ with one-pass optimized \kt axes was found to be the single most discriminating observable over a large range of signal efficiency.
Combining all observables into a multivariate discriminant, indicated a small improvement relative to just $\tau_2 / \tau_1$.

The observables were compared in data and in the simulations, in both dijet and W+jet topologies.
Selecting these topologies provided complementary samples for the jet \pt range,
and for the background composition of light-quark- and gluon-initiated jets.
Reasonable agreement was found.
In general, the \HERWIG{++} and \PYTHIA~8 generators provide better modeling
of jet substructure observables than \PYTHIA~6.
A lepton+jets $\ttbar$ sample was used to select W jets in data, and this was compared to simulation.
In this sample, we also demonstrated discrimination of the jet charge observable in data with $\PWp$ jets and $\PWm$ jets, and we studied
the performance of the W-jet tagging algorithm for a specific set of selections.
The efficiency and mistagging rate were obtained as a function of \pt.
For a typical working point, an efficiency of 65\% and a background rejection of 96\% is achieved at $\pt = 500$\GeV.
The mistagging rate for a broad range of \pt agrees reasonably with simulation.
Finally, a method using the $\ttbar$ sample was outlined for
determining data-to-simulation scale factors for correcting differences between
data and simulation of the $\tau_2 / \tau_1$ selection, the mass scales, and the resolution.

The methods introduced in this paper are directly applicable for identifying other massive objects that decay to hadrons.
For identifying Z jets, the only difference is that the jet mass window is slightly higher.
For identifying highly boosted Higgs bosons decaying to bottom quarks, the performance of these observables should be similar.
An additional discriminating variable for Higgs that is not addressed in this study,
is the possible tagging of b-jets.
We leave to future studies the optimization of the method for the boosted Higgs bosons.

\begin{acknowledgments} \hyphenation{Bundes-ministerium Forschungs-gemeinschaft Forschungs-zentren} We congratulate our colleagues in the CERN accelerator departments for the excellent performance of the LHC and thank the technical and administrative staffs at CERN and at other CMS institutes for their contributions to the success of the CMS effort. In addition, we gratefully acknowledge the computing centres and personnel of the Worldwide LHC Computing Grid for delivering so effectively the computing infrastructure essential to our analyses. Finally, we acknowledge the enduring support for the construction and operation of the LHC and the CMS detector provided by the following funding agencies: the Austrian Federal Ministry of Science, Research and Economy and the Austrian Science Fund; the Belgian Fonds de la Recherche Scientifique, and Fonds voor Wetenschappelijk Onderzoek; the Brazilian Funding Agencies (CNPq, CAPES, FAPERJ, and FAPESP); the Bulgarian Ministry of Education and Science; CERN; the Chinese Academy of Sciences, Ministry of Science and Technology, and National Natural Science Foundation of China; the Colombian Funding Agency (COLCIENCIAS); the Croatian Ministry of Science, Education and Sport, and the Croatian Science Foundation; the Research Promotion Foundation, Cyprus; the Ministry of Education and Research, Estonian Research Council via IUT23-4 and IUT23-6 and European Regional Development Fund, Estonia; the Academy of Finland, Finnish Ministry of Education and Culture, and Helsinki Institute of Physics; the Institut National de Physique Nucl\'eaire et de Physique des Particules~/~CNRS, and Commissariat \`a l'\'Energie Atomique et aux \'Energies Alternatives~/~CEA, France; the Bundesministerium f\"ur Bildung und Forschung, Deutsche Forschungsgemeinschaft, and Helmholtz-Gemeinschaft Deutscher Forschungszentren, Germany; the General Secretariat for Research and Technology, Greece; the National Scientific Research Foundation, and National Innovation Office, Hungary; the Department of Atomic Energy and the Department of Science and Technology, India; the Institute for Studies in Theoretical Physics and Mathematics, Iran; the Science Foundation, Ireland; the Istituto Nazionale di Fisica Nucleare, Italy; the Korean Ministry of Education, Science and Technology and the World Class University program of NRF, Republic of Korea; the Lithuanian Academy of Sciences; the Ministry of Education, and University of Malaya (Malaysia); the Mexican Funding Agencies (CINVESTAV, CONACYT, SEP, and UASLP-FAI); the Ministry of Business, Innovation and Employment, New Zealand; the Pakistan Atomic Energy Commission; the Ministry of Science and Higher Education and the National Science Centre, Poland; the Funda\c{c}\~ao para a Ci\^encia e a Tecnologia, Portugal; JINR, Dubna; the Ministry of Education and Science of the Russian Federation, the Federal Agency of Atomic Energy of the Russian Federation, Russian Academy of Sciences, and the Russian Foundation for Basic Research; the Ministry of Education, Science and Technological Development of Serbia; the Secretar\'{\i}a de Estado de Investigaci\'on, Desarrollo e Innovaci\'on and Programa Consolider-Ingenio 2010, Spain; the Swiss Funding Agencies (ETH Board, ETH Zurich, PSI, SNF, UniZH, Canton Zurich, and SER); the Ministry of Science and Technology, Taipei; the Thailand Center of Excellence in Physics, the Institute for the Promotion of Teaching Science and Technology of Thailand, Special Task Force for Activating Research and the National Science and Technology Development Agency of Thailand; the Scientific and Technical Research Council of Turkey, and Turkish Atomic Energy Authority; the National Academy of Sciences of Ukraine, and State Fund for Fundamental Researches, Ukraine; the Science and Technology Facilities Council, UK; the US Department of Energy, and the US National Science Foundation.

Individuals have received support from the Marie-Curie programme and the European Research Council and EPLANET (European Union); the Leventis Foundation; the A. P. Sloan Foundation; the Alexander von Humboldt Foundation; the Belgian Federal Science Policy Office; the Fonds pour la Formation \`a la Recherche dans l'Industrie et dans l'Agriculture (FRIA-Belgium); the Agentschap voor Innovatie door Wetenschap en Technologie (IWT-Belgium); the Ministry of Education, Youth and Sports (MEYS) of the Czech Republic; the Council of Science and Industrial Research, India; the HOMING PLUS programme of Foundation for Polish Science, cofinanced from European Union, Regional Development Fund; the Compagnia di San Paolo (Torino); the Consorzio per la Fisica (Trieste); MIUR project 20108T4XTM (Italy); the Thalis and Aristeia programmes cofinanced by EU-ESF and the Greek NSRF; and the National Priorities Research Program by Qatar National Research Fund.
\end{acknowledgments}
\bibliography{auto_generated}   
\cleardoublepage \appendix\section{The CMS Collaboration \label{app:collab}}\begin{sloppypar}\hyphenpenalty=5000\widowpenalty=500\clubpenalty=5000\textbf{Yerevan Physics Institute,  Yerevan,  Armenia}\\*[0pt]
V.~Khachatryan, A.M.~Sirunyan, A.~Tumasyan
\vskip\cmsinstskip
\textbf{Institut f\"{u}r Hochenergiephysik der OeAW,  Wien,  Austria}\\*[0pt]
W.~Adam, T.~Bergauer, M.~Dragicevic, J.~Er\"{o}, C.~Fabjan\cmsAuthorMark{1}, M.~Friedl, R.~Fr\"{u}hwirth\cmsAuthorMark{1}, V.M.~Ghete, C.~Hartl, N.~H\"{o}rmann, J.~Hrubec, M.~Jeitler\cmsAuthorMark{1}, W.~Kiesenhofer, V.~Kn\"{u}nz, M.~Krammer\cmsAuthorMark{1}, I.~Kr\"{a}tschmer, D.~Liko, I.~Mikulec, D.~Rabady\cmsAuthorMark{2}, B.~Rahbaran, H.~Rohringer, R.~Sch\"{o}fbeck, J.~Strauss, A.~Taurok, W.~Treberer-Treberspurg, W.~Waltenberger, C.-E.~Wulz\cmsAuthorMark{1}
% ...
\vskip\cmsinstskip
% ...
1:~~Also at Vienna University of Technology, Vienna, Austria\\
2:~~Also at CERN, European Organization for Nuclear Research, Geneva, Switzerland\\
% ...

\end{sloppypar}
\end{document}